\documentclass[structabstract,dvipsnames]{aa}
\pdfoutput=1

\ifx\pdfoutput\undefined
\usepackage{graphicx}
\else
\usepackage[pdftex]{graphicx}
\usepackage[pdftex,pdfauthor={J.~Threlfall},citecolor=blue,linkcolor=blue, linktocpage=true,colorlinks=true]{hyperref}
\usepackage{epstopdf}
\fi 

\usepackage{natbib}
\bibpunct{(}{)}{;}{a}{}{,} 
\usepackage{txfonts}
\usepackage[usenames]{color}
\usepackage{amssymb}	
\usepackage{aas_macros}	
\usepackage{subfig}	
\usepackage{fixltx2e}	
\usepackage{mathrsfs}
\usepackage{hypcap}
\usepackage{multirow}

\graphicspath{{figs/pdffigs/}}

 \DeclareGraphicsExtensions{.jpg,.pdf,.png}

\newcommand{\beq}{ \begin{eqnarray} }
\newcommand{\eeq}{ \end{eqnarray} }

\def\A{{Alfv\'en}}

\def\Aic{Alfv\'enic}

\date{}			

\begin{document}
\title{First comparison of wave observations from CoMP and AIA/SDO}
\author{J.~Threlfall\inst{\ref{inst1}} \and I.~De Moortel\inst{\ref{inst1}} \and S.~W.~McIntosh\inst{\ref{inst2}} \and C.~Bethge\inst{\ref{inst2}}}
\institute{School of Mathematics and Statistics, University of St Andrews, St Andrews, Fife, KY16 9SS, U.K. \email{jamest@mcs.st-and.ac.uk;ineke@mcs.st-and.ac.uk}\label{inst1}
\and High Altitude Observatory, National Center for Atmospheric Research, PO Box 3000, Boulder, CO 80307, USA \email{mscott@ucar.edu;bethge@ucar.edu}\label{inst2}}

\abstract
{Waves have long been thought to contribute to the heating of the solar corona and the generation of the solar wind. Recent observations have demonstrated evidence of quasi-periodic longitudinal disturbances and ubiquitous transverse wave propagation in many different coronal environments.}
{This paper investigates signatures of different types of oscillatory behaviour, both above the solar limb and on-disk, by comparing findings from the Coronal Multi-channel Polarimeter (CoMP) and the Atmospheric Imaging Assembly (AIA) on board the Solar Dynamics Observatory (SDO) for the same active region.}
{We study both transverse and longitudinal motion by comparing and contrasting time-distance images of parallel and perpendicular cuts along/across active region fan loops.  Comparisons between parallel space-time diagram features in CoMP Doppler velocity and transverse oscillations in AIA images are made, together with space-time analysis of propagating quasi-periodic intensity features seen near the base of loops in AIA.}
{Signatures of transverse motions are observed along the same magnetic structure using CoMP Doppler velocity ($v_{\rm{phase}}=600\rightarrow750\,\rm{km~s}^{-1}$, $P=3\rightarrow6\,\rm{mins}$) and in AIA/SDO above the limb ($P=3\rightarrow8\,\rm{mins}$). Quasi-periodic intensity features ($v_{\rm{phase}}=100\rightarrow200\,\rm{km~s}^{-1}$, $P=6\rightarrow11\,\rm{mins}$) also travel along the base of the same structure. On the disk, signatures of both transverse and longitudinal intensity features were observed by AIA, and both show similar properties to signatures found along structures anchored in the same active region three days earlier above the limb. Correlated features are recovered by space-time analysis of neighbouring tracks over perpendicular distances of $\lesssim2.6\,\rm{Mm}$.}
{}

\maketitle
\newcommand{\angstrom}{\mbox{\normalfont\AA}}

\section{Introduction}\label{sec:intro}

Over the past 15 years an increasing body of evidence has demonstrated that wave propagation (and dissipation) occurs in the solar corona. Such evidence provides a way to assess the potential contribution of waves both to the heating of the solar atmosphere and the generation of the solar wind \citep[see e.g.][for a recent review]{review:ParnellDeMoortel2012}. In addition, MHD waves may also be used to infer estimates of local plasma parameters that are otherwise difficult to measure directly. These seismological tools rely on accurate identification of MHD wave-modes and their properties  \citep[see reviews of][]{review:Roberts2000,review:NakariakovVerwichte2005,review:Banerjeeetal2007,review:DeMoortelNakariakov2012}.  

Evidence of impulsively generated (e.g. flare-induced) transverse oscillatory behaviour has become relatively common, beginning with examples from the Transition Region and Coronal Explorer \citep[TRACE, e.g.][]{paper:Aschwandenetal1999, paper:Nakariakovetal1999} right through to the present \citep[e.g.][]{paper:WhiteVerwichte2012}. Meanwhile, recent advances in imaging technology have revealed footpoint driven transverse waves propagating along many coronal structures, for example, along spicules \citep{paper:DePontieuetal2007}, X-ray jets \citep{paper:Cirtainetal2007} and prominences \citep{paper:Okamotoetal2007}. Footpoint-driven transverse waves have also been observed along coronal loops \citep[][]{paper:Tomczyketal2007,paper:TomczykMcIntosh2009,paper:McIntoshetal2011} and are now thought to be ubiquitous; see \citet{review:Mathioudakisetal2012} for a recent review.

\begin{figure*}[t]
 \centering\capstart
 \subfloat[AIA $193\angstrom$ intensity]{\label{subfig:limbAIA}{\resizebox{0.415\textwidth}{!}{\includegraphics{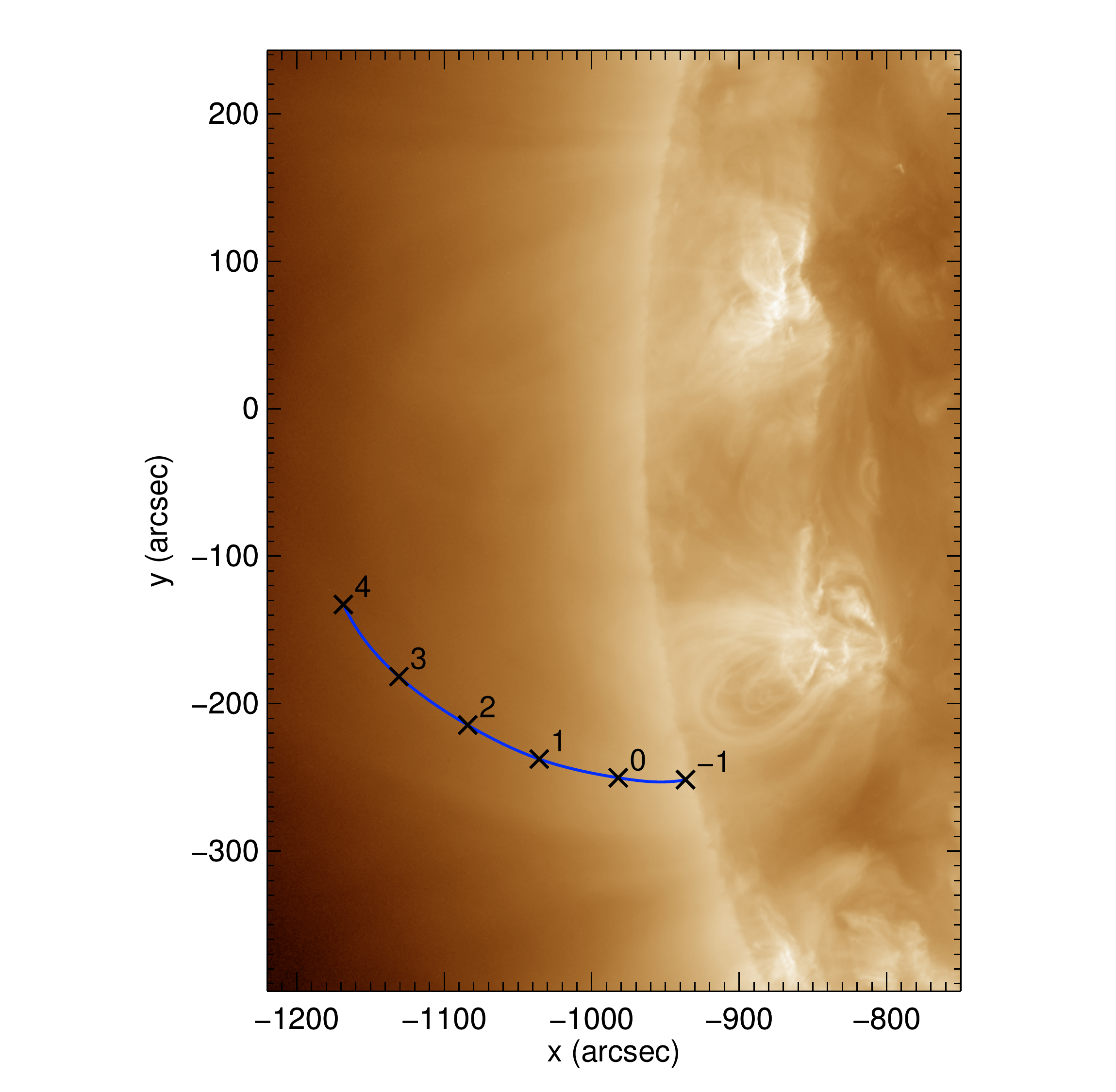}}}}
\subfloat[Enhanced AIA $193\angstrom$]{\label{subfig:limbAIAUSM}{\resizebox{0.57\textwidth}{!}{\includegraphics{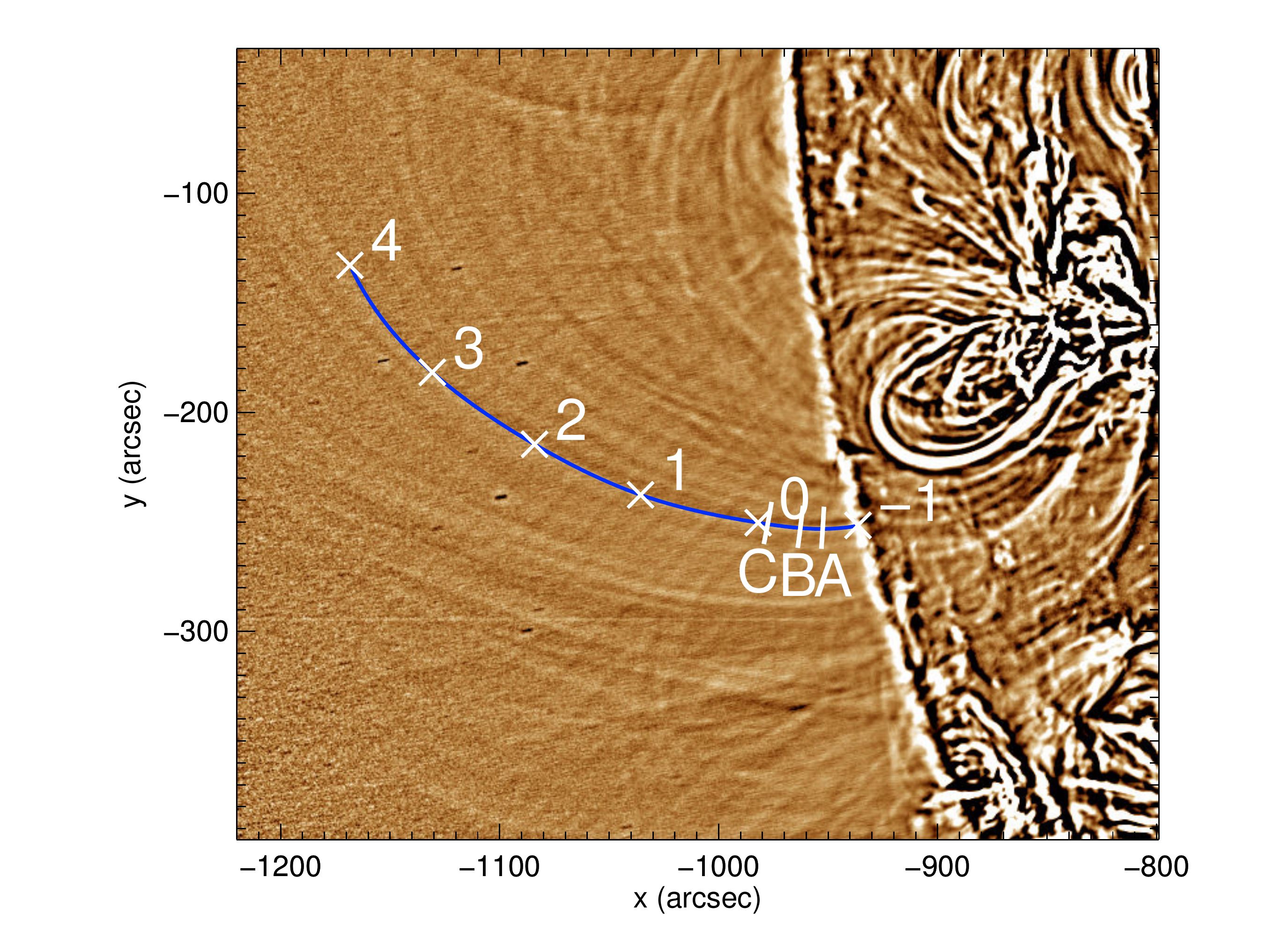}}}}
\caption{AIA $193\angstrom$ intensity on the eastern solar limb at 17:10UT, 11 April 2012. In order to bring out the magnetic structures present above the limb, the pure intensity images \protect\subref{subfig:limbAIA} are summed over a period of an hour, then unsharp masked in \protect\subref{subfig:limbAIAUSM}. The track studied is highlighted in blue, and several perpendicular cross sections were taken at {{locations A-C}}.}
 \label{fig:11Apr_refim}
\end{figure*}

The nature of these ubiquitous transverse wave observations, particularly those seen by the Coronal Multi-channel Polarimeter \citep[CoMP,][]{paper:Tomczyketal2008} has been a matter of debate. With speeds much greater than the local sound speed, propagation along magnetic field lines and no evidence of significant intensity fluctuations (implying largely incompressible motion), these waves were originally identified as (shear) {\A} waves \citep{paper:Tomczyketal2007,paper:TomczykMcIntosh2009}. Subsequent theoretical studies \citep[e.g.][]{paper:VanDoorsselaereetal2008} have suggested an interpretation of these observations using propagating fast magnetoacoustic (MA) kink waves.  However, in a 3D geometry, for loops which are (even weakly) stratified in density in the transverse direction, 3D MHD simulations have shown that generic transverse footpoint displacements lead to coupled kink-Alfv\'en modes \citep[][]{paper:Pascoeetal2010,paper:Pascoeetal2012,paper:Pascoeetal2013}. The term ``{\Aic}'' is often used to describe these inherently coupled waves. For a more comprehensive discussion we refer the interested reader to the Supplementary Material accompanying \cite{paper:McIntoshetal2011} and \citet{paper:Goossensetal2012a}. Furthermore, analysing observations by the Swedish Solar Telescope, \citet{paper:DePontieuetal2012} suggest that, in addition to the transverse motions described above, type II spicules {\it simultaneously} undergo (ubiquitous) torsional motions as well as supporting field-aligned flows. 

The broadening of spectral emission lines as a function of height in the corona has often been interpreted as a signature of {\A} waves, responsible for unresolved motions occurring with a component along the line of sight \citep[see e.g.][]{paper:Hassleretal1990,paper:HasslerMoran1994,paper:Banerjeeetal1998,paper:Erdelyietal1998,paper:Doyleetal1998,paper:Moran2001,paper:Moran2003,paper:Singhetal2006}. More recently, \citet{paper:Jessetal2009} used spectral emission line broadening (unaccompanied by periodicities in intensity or line of sight velocity) to infer the presence of torsional {\A} waves, travelling from the chromosphere to the corona along a magnetic flux tube. The specific amount by which emission lines broaden has also been used to infer {\A} wave damping above certain heights in the corona \citep[see e.g.][and references therein]{paper:BemporadAbbo2012}. 

Several observations also demonstrate small amplitude quasi-periodic intensity disturbances propagating upward from the base of coronal loops \citep[see e.g.][for a comprehensive review]{review:DeMoortel2009}. These propagating coronal disturbances (PDs/PCDs) are commonly seen as an alternating series of ridges within space-time diagrams of intensity. These ridges typically only exist over the lowest $30\mathord{-}50\,\rm{Mm}$ of these loops, with small amplitudes (a few percent above the background), speeds close to the local sound speed and periods from $2\mathord{-}10\,\rm{mins}$. Such properties have often lead to their classification as coronal signatures of slow magnetoacoustic waves \citep[see][and references therein]{paper:Banerjeeetal2011}. Recent spectroscopic observations have begun to cast doubt over such a definitive classification. Examination of periodic features occurring in phase in intensity, Doppler velocity, line width and line asymmetry, observed by the Extreme-Ultraviolet Imaging Spectrometer \citep[EIS/Hinode][]{paper:EIS} has given strength to an alternative interpretation; \citet{paper:DePontieuMcIntosh2010} and \citet{paper:Tianetal2011a} have shown that intermittent upflows produce coherent behaviour in all four line parameters. However, slow magnetoacoustic waves may also be responsible for these signatures \citep[see e.g.][]{paper:Verwichteetal2010,inproc:Wangetal2012}. Definitive wave/flow classification of observations may be further complicated by a number of factors \citep[for example, the local environment was shown to play a key role by][who surveyed many examples of quasi-periodic behaviour, and recovered a temperature dependence in velocities of features above sunspots, thus favouring a wave interpretation in those cases]{paper:Kiddieetal2012}. Recently, efforts have been made to use co-spatial and co-temporal spectroscopic observations from EIS/Hinode and image sequences from the Atmospheric Imaging Assembly \citep[AIA;][]{paper:AIA} on board the Solar Dynamics Observatory \citep[SDO;][]{paper:SDO} to study these disturbances in more detail.

Both spectral and imaging data from the two instruments are used by \citet{paper:Tianetal2011c} to imply that some intensity oscillations are upflows which may be responsible for blueward asymmetries in emission line profiles. \citet{paper:KrishnaPrasadetal2012a} identify both long and short period oscillations (of which the former are only seen using spectroscopy) and favour a slow magnetoacoustic wave interpretation. \citet{paper:McIntoshetal2012} used both instruments to link the observed motions to the mass transport cycle of the corona and chromosphere. 

Of particular interest for the present study is \citet{paper:Tianetal2011b}, who identify outflows (with speeds of ${\mathord{\sim}}120\,\rm{km~s}^{-1}$) along coronal plumes upon which transverse motions are also reported. A statistical survey of EIS findings was performed in \citet{paper:Tianetal2012}, which identifies two specific types of common oscillatory behaviour; one type is associated with features acting in phase in all line parameters at the base of loop footpoints, while another occurs at much larger heights, primarily in Doppler, with typical periods of $3-6\,\rm{mins}$. Thus spectroscopic measurements, in conjunction with imaging methods, are able to identify both longitudinal and transverse motions, even on the same structure. At present, many different authors continue to reach different conclusions over a wave or flow classification of observations; both categories are capable of producing signatures which may be recovered using the other, and hence care must be taken when using such observations as the basis for seismological models.

The primary objective of this work is to investigate whether any of the signatures of wave-like behaviour previously identified separately by CoMP and AIA/SDO may be seen on the same magnetic structures at the same time. By comparing the properties of these signatures over the course of several days, we will look to establish how off-limb observations using CoMP and off-limb/on-disk features seen by AIA are related. Following an outline of our observational data preparation (Sec.~\ref{sec:dataprep}), we present an overview of the techniques used to recover multiple distinct types of wave-like motions (Sec.~\ref{sec:meth}). Our results begin with an examination of propagating transverse oscillations along fan loops observed on 11 April 2012 in CoMP Doppler velocity and line width (Sec.~\ref{sec:comp11}). Studying the same set of fan loops with AIA reveals evidence that the loops may be supporting both transverse and longitudinal waves (Sec.~\ref{sec:AIA11}). Section~\ref{sec:AIA14} compares properties of motions found along similar fan loops three days apart. A discussion of the results can be found in Sec.~\ref{sec:disc}, before conclusions are presented in Sec.~\ref{sec:conc}.

\section{Observations: data preparation}\label{sec:dataprep}

A pair of active regions are visible on the eastern solar limb, close to the equator on 11 April 2012 (see Fig.~\ref{subfig:limbAIA}). Images obtained from AIA and the Helioseismic and Magnetic Imager \citep[HMI;][]{paper:HMI} on board SDO, show that neither is associated with a sunspot during their transit over the solar disk. Further images from both AIA and the Extreme Ultraviolet Imager \citep[EUVI/STEREO-B][]{paper:EUVI,paper:STEREOintro} show a range of both open fan loop structures and loops which link both active regions and extend to large heights above the solar limb (see Fig.~\ref{subfig:limbAIAUSM}).

Our initial investigation starts at 17:11:53UT on 11 April 2012, when a continuous $2\rm{hr}$ imaging sequence was taken by CoMP, a ground-based instrument {{(located at the Mauna Loa Solar Observatory, Hawaii)}} with a wide field-of-view which observes the solar corona from $1.05$ to $1.35$ solar radii, at a spatial sampling of ${{4.46\,\rm{arcsec/pixel}}}$ and with a cadence of $30\,\,\rm{s}$. Line-centred intensity, Doppler velocity and line width are based on measurements of the FeXIII coronal $1074.7\,\rm{nm}$ emission line.
The data has been corrected for an east-west trend in the velocities, and the zero-point of the wavelength scale has been redefined under the assumption that the mean Doppler shift over the field of view is zero; finally, all frames have been coaligned using a cross-correlation technique.

Due to an SDO calibration event coinciding with the time range available from CoMP on this date, the closest available AIA data window begins at 15:59:37UT, lasting for $70{\,\rm{mins}}$. We obtained $12\,\rm{s}$ cadence data from the $193\angstrom$ and $171\angstrom$ AIA passbands; both AIA datasets have been cleaned and coaligned using the SolarSoft IDL \verb|AIA_prep| command, before being derotated. Finally, three consecutive AIA frames were summed to boost signal-to-noise, yielding a $36\,\rm{s}$ cadence. Both CoMP and AIA data obtained on 11 April 2012 were cropped to focus on a small region at the equator on the eastern solar limb, containing the two active regions (see Fig.~\ref{subfig:limbAIA}). In order to study the long term evolution of these active regions (over several days), data was also obtained at 15.33.08UT on 14 April 2012, lasting approximately 3 hours.

\begin{figure}[t]
  \centering \capstart
  \resizebox{0.49\textwidth}{!}{\includegraphics{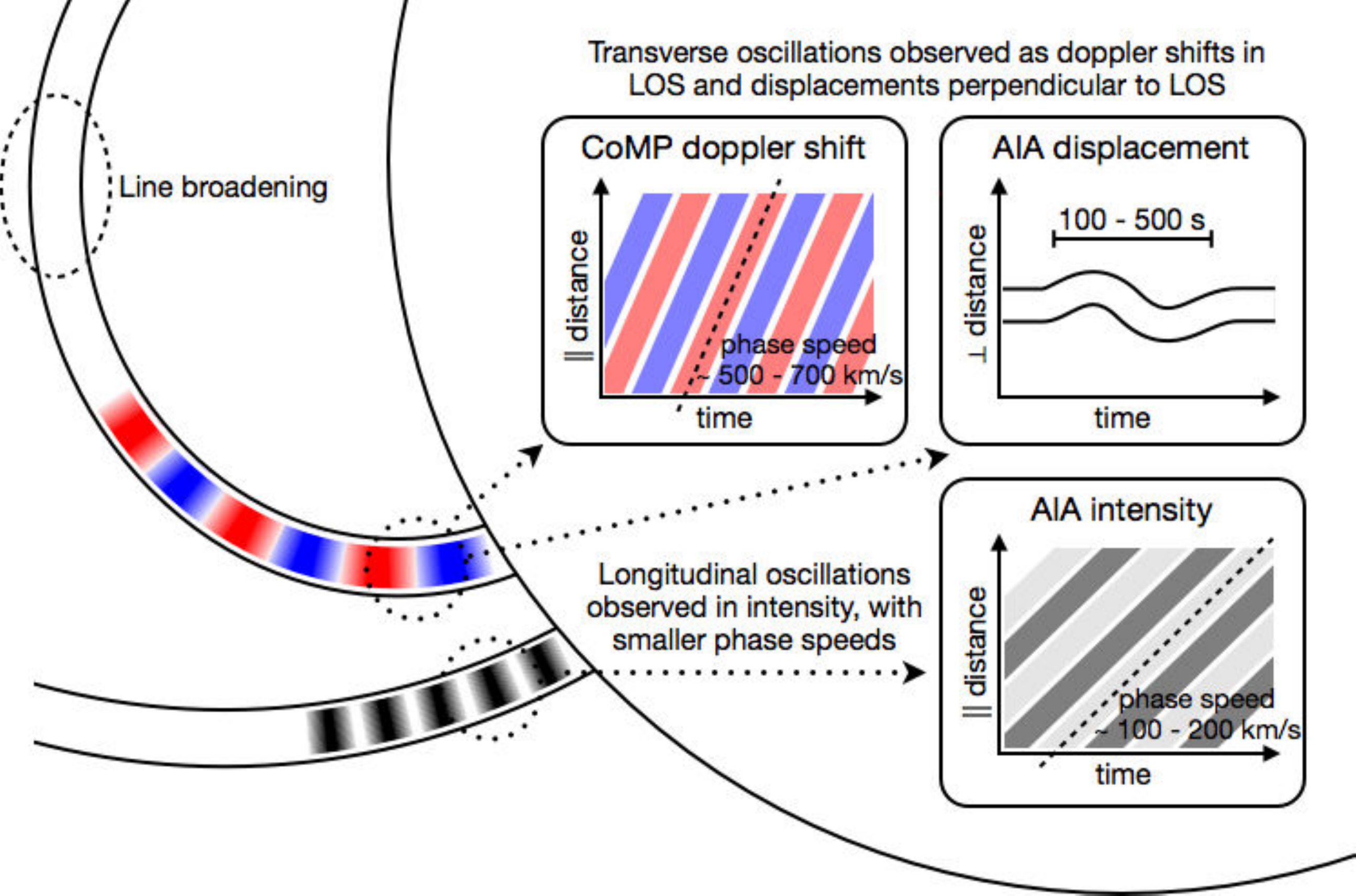}}
  \caption{Illustration of the methods used to identify propagating wave-like phenomena along the same magnetic structure, using different instruments. Red/blue indicates Doppler shift oscillations whereas black/white corresponds to intensity variations}
  \label{fig:cartoon}
 \end{figure}

\section{Methods}\label{sec:meth}

Several methods are used in this study to identify specific types of oscillatory/wave behaviour present. 

\begin{figure*}[t]
  \centering\capstart
  \subfloat[CoMP enhanced intensity]{\label{subfig:11Apr_CoMP_I}\resizebox{0.33\textwidth}{!}{\includegraphics{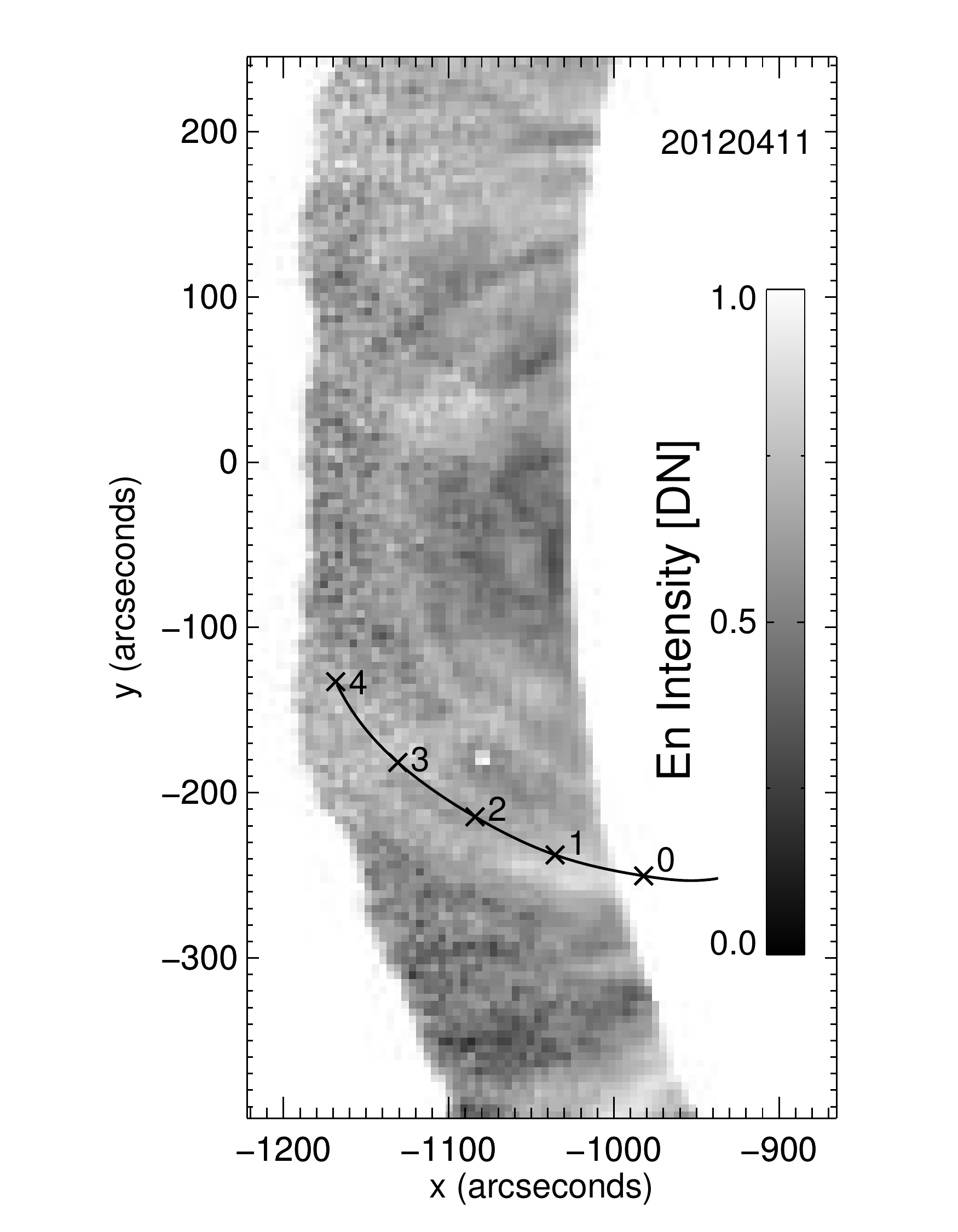}}}
  \subfloat[CoMP line width]{\label{subfig:11Apr_CoMP_LW}\resizebox{0.33\textwidth}{!}{\includegraphics{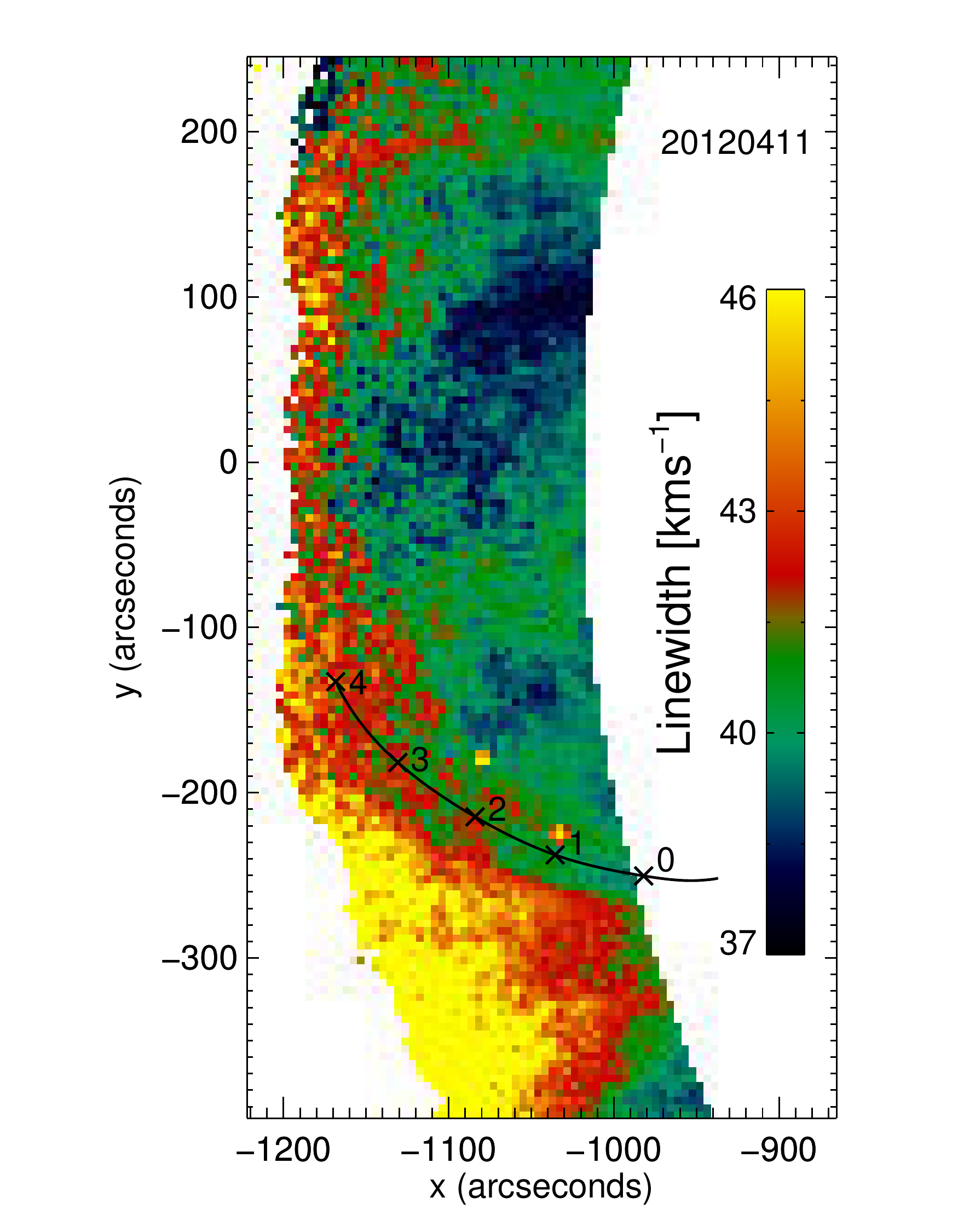}}}
  \subfloat[CoMP Doppler velocity]{\label{subfig:11Apr_CoMP_DV}\resizebox{0.33\textwidth}{!}{\includegraphics{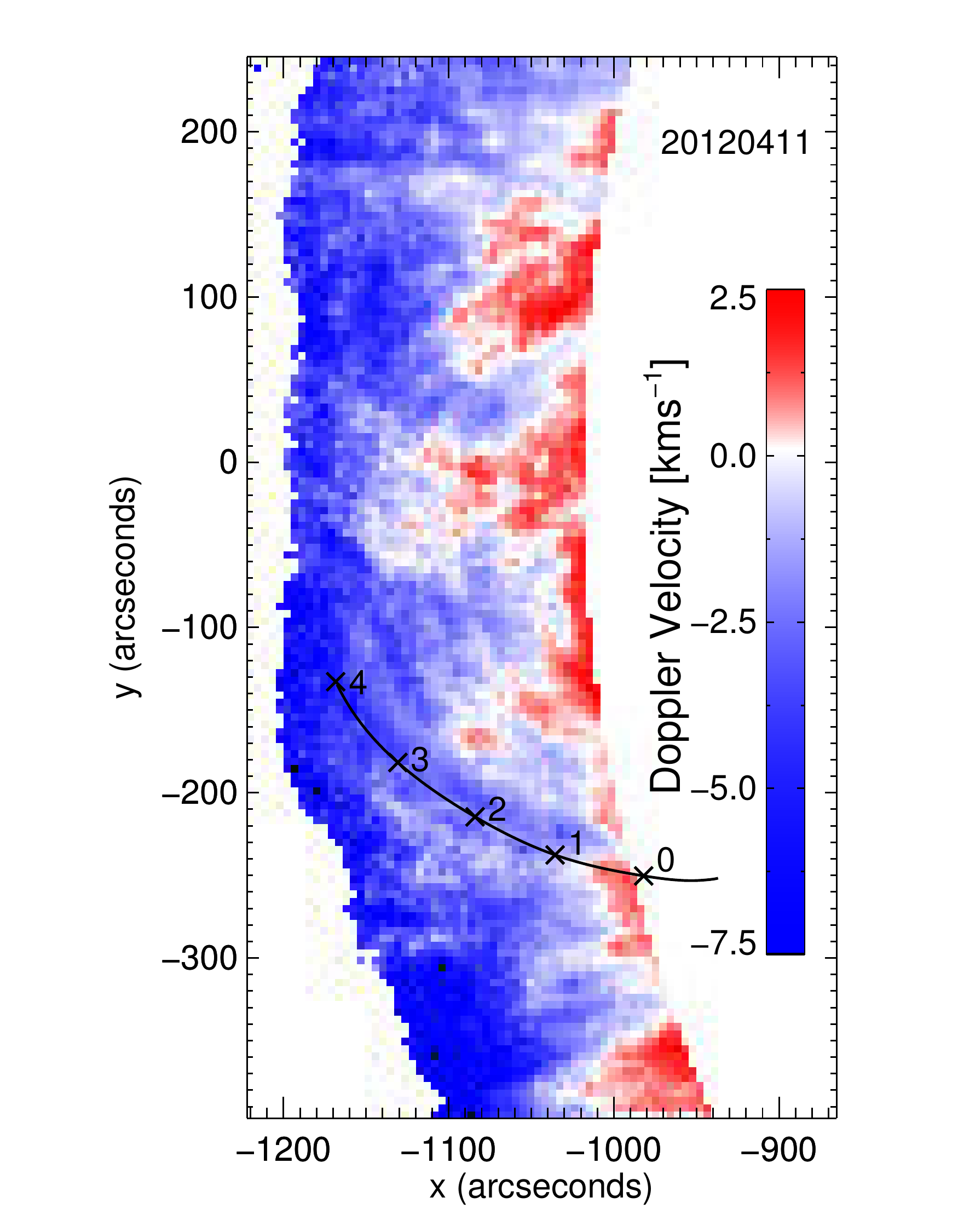}}}
 \caption{Typical snapshots of the quantities observed using CoMP; features seen in radially enhanced intensity \protect\subref{subfig:11Apr_CoMP_I}, line width \protect\subref{subfig:11Apr_CoMP_LW} and Doppler velocity \protect\subref{subfig:11Apr_CoMP_DV} in the region of interest on the solar limb (seen in Fig.~\ref{subfig:limbAIA} using AIA). {{Features seen in red in \protect\subref{subfig:11Apr_CoMP_DV} have been red-shifted, i.e. motion away from the observer (positive velocities) whereas blue corresponds to motion towards the observer (negative velocities).}}}
 \label{fig:11Apr_CoMPtracks}
 \end{figure*}

In the field-of-view (FOV) of each instrument, we begin by defining an arc along a feature we wish to study (the long blue arc seen in Fig~\ref{subfig:limbAIAUSM}). The coordinates of the arc are resampled to be equally spaced; for simplicity, the spacing is chosen to match the pixel-separation distance of the instruments. Thus, an arc studied in CoMP data is sampled every ${{4.46\,\rm{arcsecs}}}$ ($3.24\,\rm{Mm}$) along its entire length, while an arc in AIA is sampled every $0.6\,\rm{arcsecs}$ ($0.44\,\rm{Mm}$). In addition, we store data in a ${\mathord{\sim}}20\,\rm{Mm}$ perpendicular cut to the arc for each position along the arc. The data in each perpendicular cut is also spaced by the instrumental pixel separation distance. Thus, we build up a grid of perpendicular cuts, centred on each arc, with a cut at every position along the structure. This technique allows us to study longitudinal and transverse behaviour simultaneously.

\begin{figure}[t]
  \centering \capstart
  \resizebox{0.48\textwidth}{!}{\includegraphics{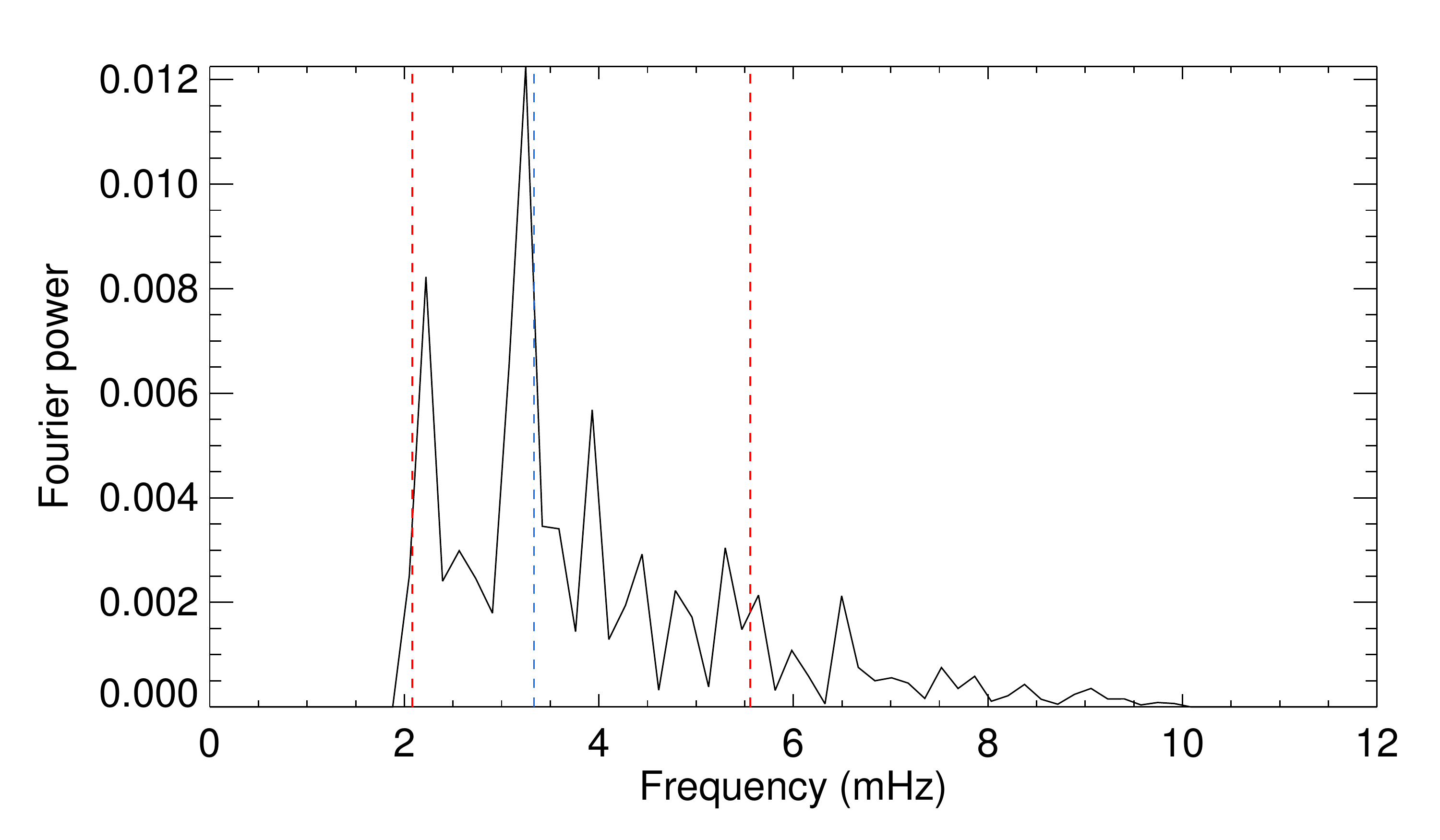}}
  \caption{Distribution of Fourier power averaged along the chosen arc in CoMP Doppler velocity (Fig.~\ref{subfig:11Apr_CoMP_DV}). Several specific frequencies are highlighted with dashed lines; from left to right, these correspond to $8$ (red), $5$ (blue) and $3\,\rm{min}$ (red) periods.} 
  \label{fig:11Apr_CoMPFFT}
 \end{figure}
 
 \begin{figure*}[t]
  \centering \capstart
  \resizebox{0.98\textwidth}{!}{\includegraphics{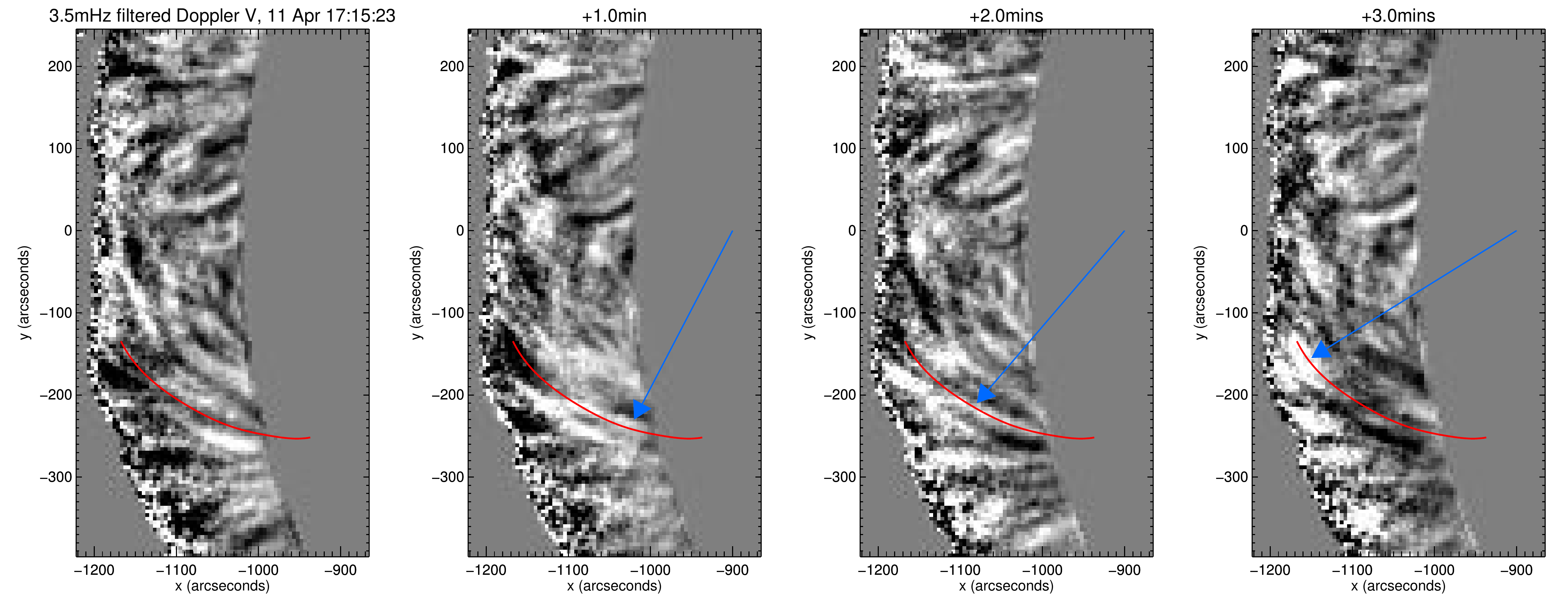}}
  \caption{Example of a propagating feature in $5\,\rm{min}$ filtered CoMP Doppler velocity images along the chosen track; highlighted with a blue arrow, coherent features travel outwards from the limb (at around $0.5\,\rm{Mm\,s}^{-1}$) along the highlighted structure.} 
  \label{fig:11Apr_CoMPDopplerProp}
 \end{figure*}

The choice of whether to perform space-time analysis of longitudinal or perpendicular data is determined by the type of wave-like behaviour we hope to observe, and the instrument used. CoMP Doppler velocities reveal displacements along the line-of-sight above the solar limb. Performing a space-time analysis along a single structure can be used to estimate the phase speed and wave power of these waves. Previous studies with CoMP have shown little/no evidence of significant intensity perturbations associated with the observed Doppler shifts, implying that these waves are incompressible.

The turbulent convective motions which are thought to drive the observed transverse waves will be randomly polarised. Hence, in addition to displacements along the line-of-sight (observed as Doppler shift oscillations), it is likely there will also be a plane-of-sky component, observable as loop (intensity) displacements. However, the wave amplitudes recorded by CoMP Doppler velocity ($\lesssim2\,\rm{km\,s}^{-1}$) imply that similar displacements in the plane-of-sky would not be resolved by CoMP intensity (with the $3.24\,\rm{Mm}$ pixel width of the CoMP instrument). Perpendicular cuts across loops using high resolution imagers (including AIA) have resolved small transverse loop displacements in the plane-of-sky \citep[e.g.][]{paper:McIntoshetal2011}. Coupling the space-time analysis of perpendicular cuts across structures in AIA intensity with properties found using CoMP Doppler velocity along the same structure allows us to study two aspects of transverse motion simultaneously.

Finally, AIA has also been used to examine longitudinal quasi-periodic propagating intensity perturbations travelling along magnetic structures. Space-time analysis along the central column of our grid would highlight the presence of any intensity features propagating along our chosen arc as a series of ridges; the orientation, separation and amplitude of these ridges may again reveal the properties of these longitudinal motions.

By combining these techniques, we build up a picture of the wave behaviour occurring along and across magnetic structures in the corona; this picture is illustrated by the cartoon in Fig.~\ref{fig:cartoon}.

\section{11 April 2012: CoMP}\label{sec:comp11}

We begin our investigation with a study of the active regions off-limb on 11 April 2012. We used AIA to identify specific structures of interest, due to its high spatial resolution. To enhance the structures seen in intensity (Fig.~\ref{subfig:limbAIA}), we employed an unsharp mask edge-enhancement technique (subtracting a smoothed version of the logarithm of the image from the logarithm of the original image). This allowed us to identify an arc which closely follows a fan loop anchored in the southern active region (Fig.~\ref{subfig:limbAIAUSM}). The coordinates of this arc were then translated onto the CoMP observation, in order to compare the same feature using both instruments; the same fan loop seen by AIA in Fig.~\ref{fig:11Apr_refim} is shown in the CoMP field-of-view in Fig.~\ref{fig:11Apr_CoMPtracks}.

CoMP Doppler velocity shows continuous, footpoint-driven transverse waves throughout this region, travelling along magnetic structures \citep[as seen by][]{paper:Tomczyketal2007,paper:TomczykMcIntosh2009}. As with the earlier results from CoMP by Tomczyk and coauthors, no significant variations in intensity are associated with these motions. The Fourier spectrum of the measured Doppler velocity averaged along the arc peaks close to $3.5\,\rm{mHz}$ (see Fig.~\ref{fig:11Apr_CoMPFFT}). Filtering each image using a Gaussian filter (centred on $3.5\,\rm{mHz}$, with a width of $0.5\,\rm{mHz}$) enhances the visibility of these Doppler shifts, in order to track a series of these displacements propagating along our chosen arc (see example in Fig.~\ref{fig:11Apr_CoMPDopplerProp}). These features appear to travel much faster than estimates of the local sound speed; at the formation temperature of the $1074.7\,\rm{nm}$ Fe XIII line \citep[$T=1.6\,\rm{MK}$][]{paper:Judgeetal2006}, the local sound speed $c_s$ is approximately $192\,\rm{km\,s}^{-1}$ \citep[using $c_s\sim152\, T^{1/2}{\rm{m\,s}^{-1}}$,][]{book:Priest1982}.

 \begin{figure*}[t]
  \centering\capstart
  \subfloat[CoMP Doppler]{\label{subfig:11Apr_DVTD}\resizebox{0.48\textwidth}{!}{\includegraphics{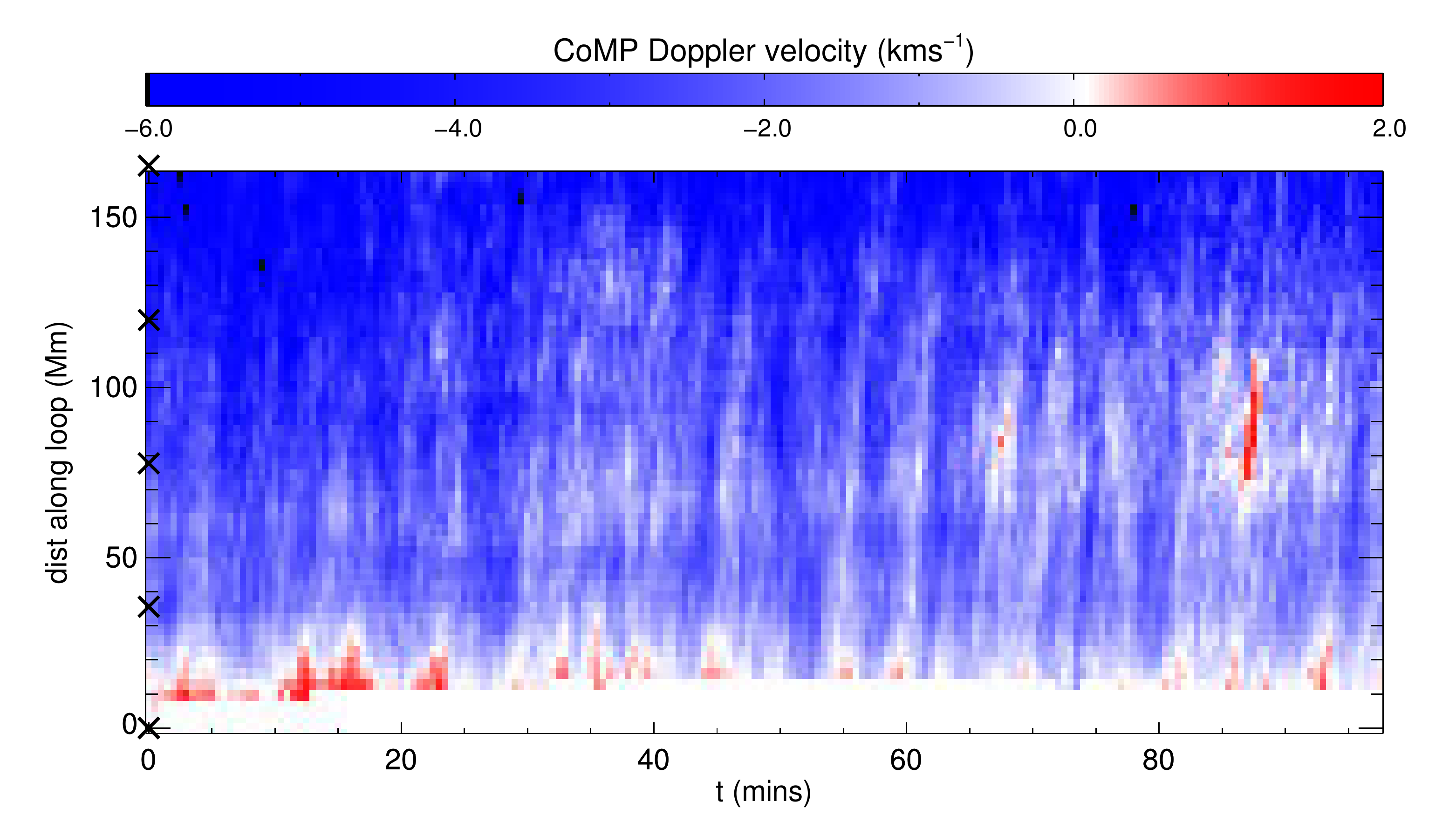}}}
  \subfloat[CoMP line width]{\label{subfig:11Apr_LWTD}\resizebox{0.48\textwidth}{!}{\includegraphics{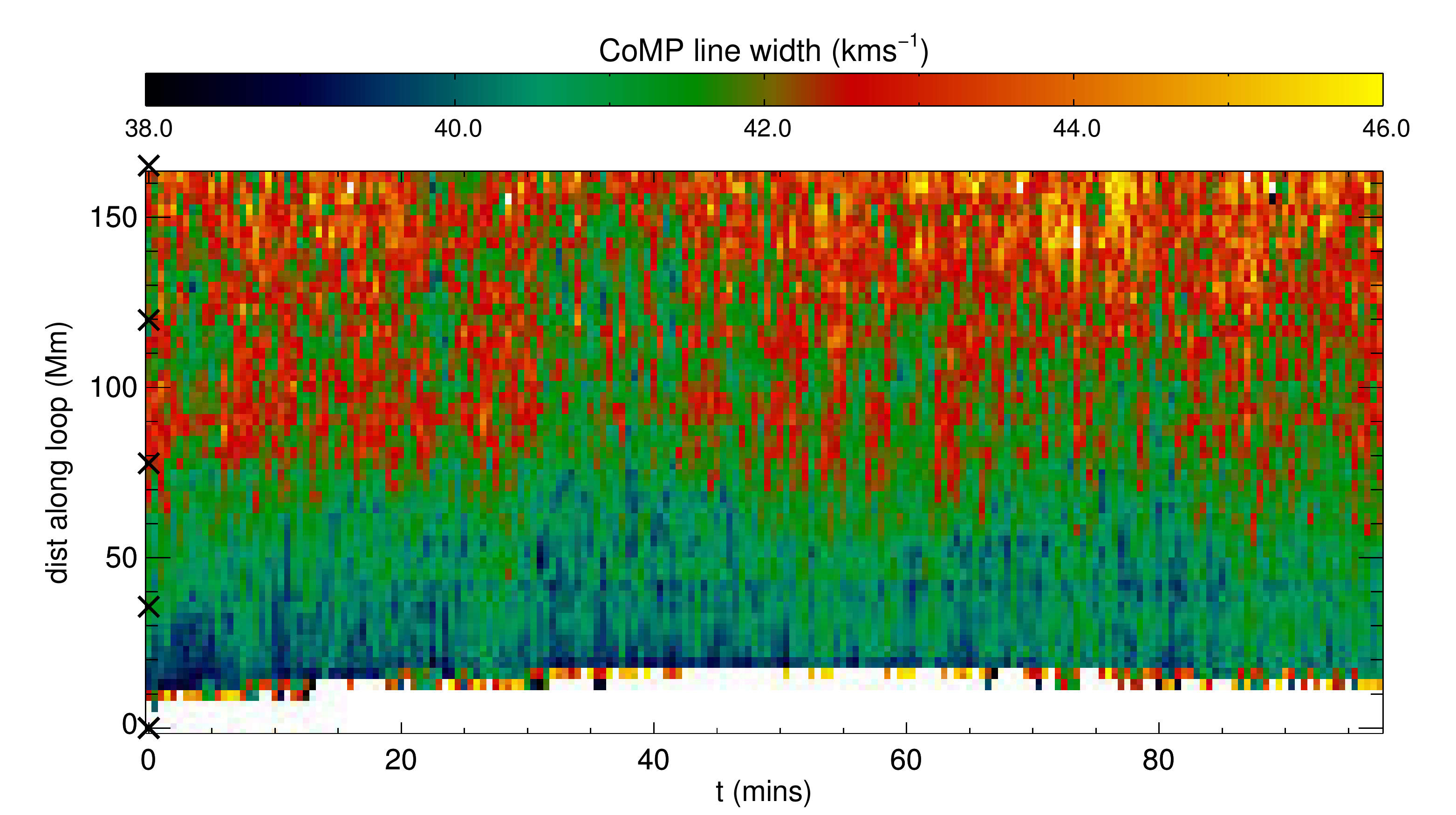}}}\\
  \subfloat[Doppler, $9$min RA subtracted, smoothed]{\label{subfig:11Apr_DVTDRAs}\resizebox{0.48\textwidth}{!}{\includegraphics{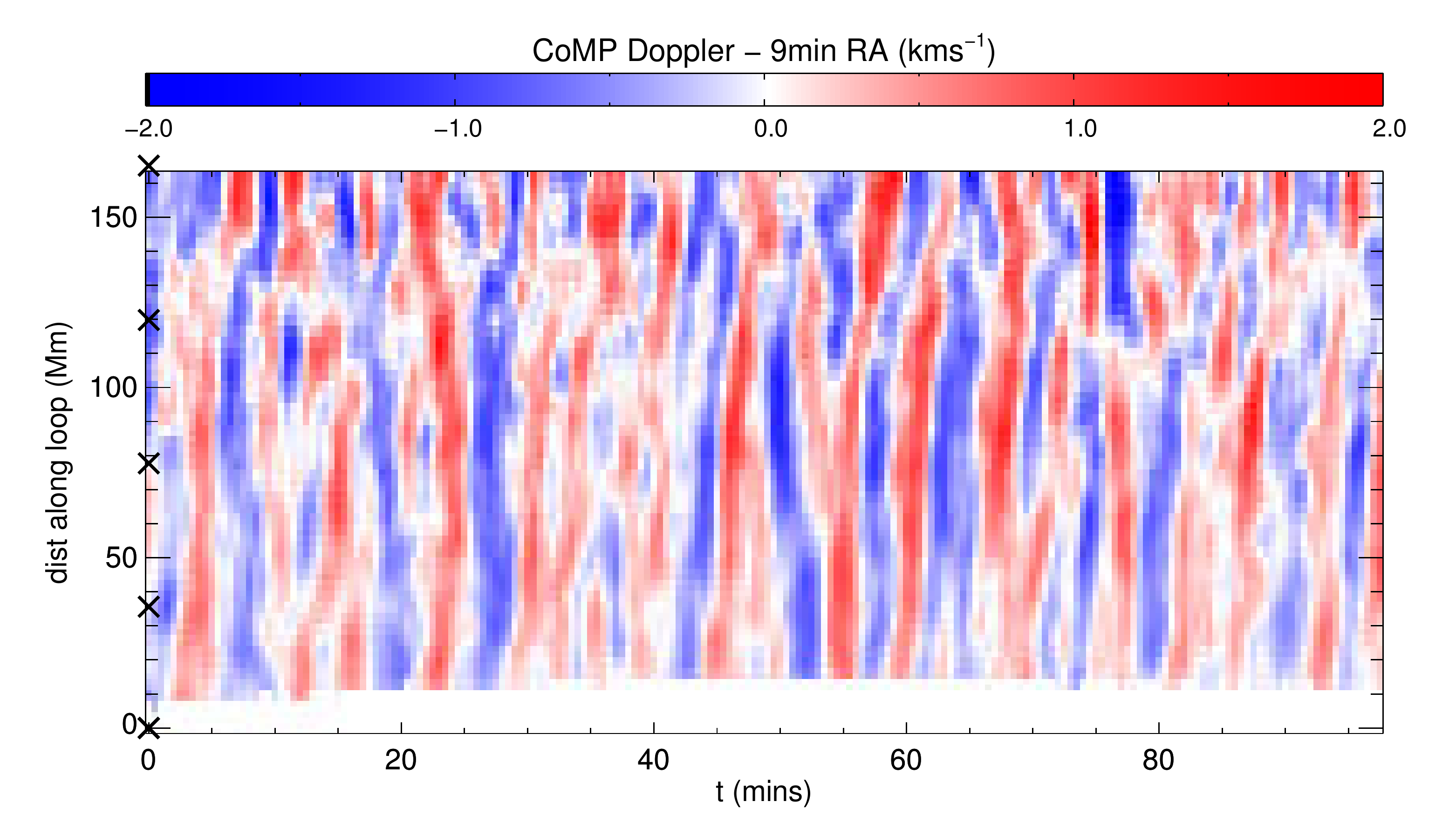}}}
  \subfloat[Line width, $9$min RA subtracted, smoothed]{\label{subfig:11Apr_LWTDRAs}\resizebox{0.48\textwidth}{!}{\includegraphics{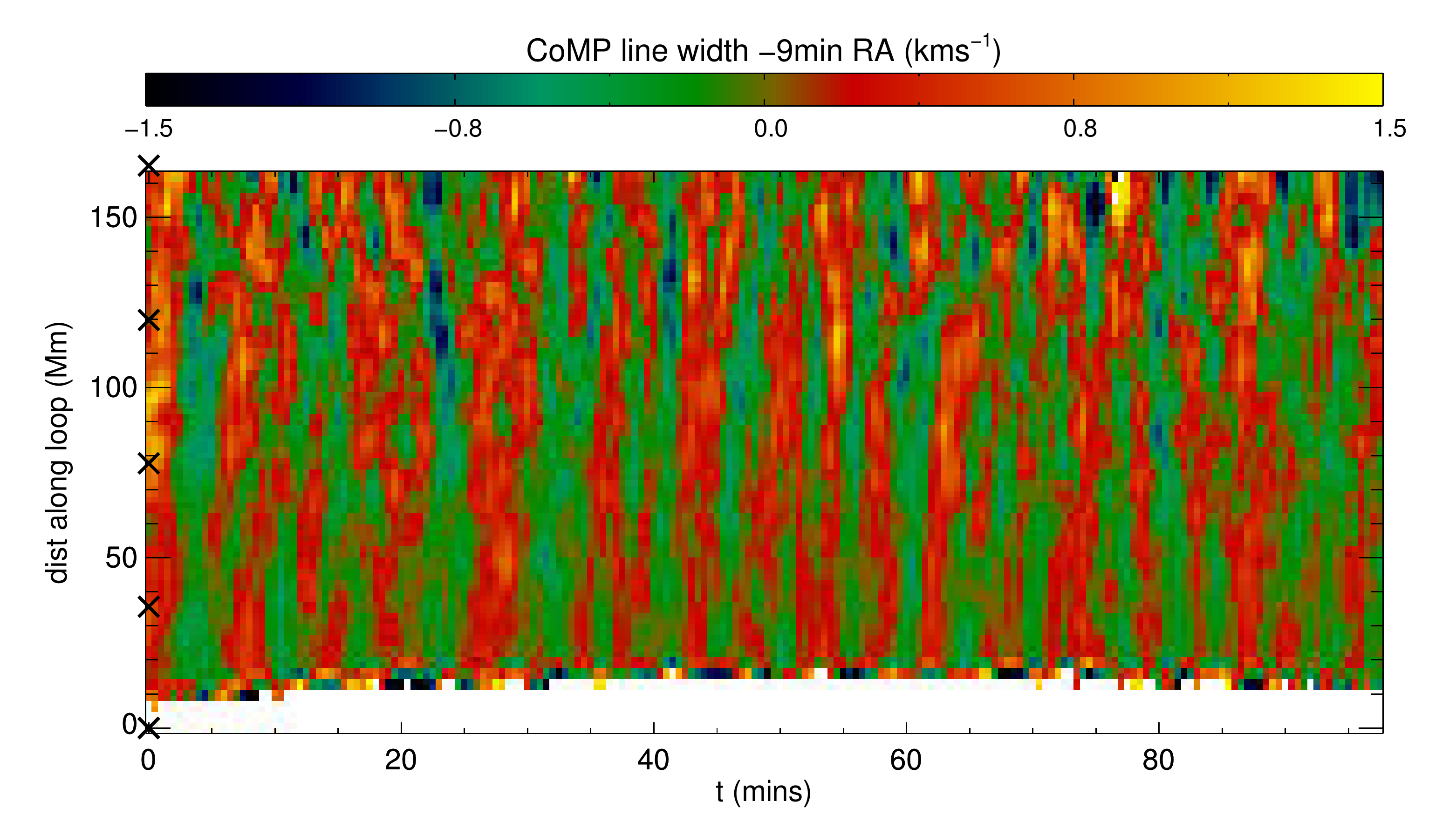}}}\\
  \caption{Doppler and line width data \protect\subref{subfig:11Apr_DVTD}-\protect\subref{subfig:11Apr_LWTD} sampled along the blue track seen in Fig.~\ref{subfig:11Apr_CoMP_I}. To enhance propagating oscillatory features, a $9\,\rm{min}$ running average is then subtracted from the Doppler~\protect\subref{subfig:11Apr_DVTDRAs} and line width~\protect\subref{subfig:11Apr_LWTDRAs} data. [NB. the white region near the bottom of the track is caused by the CoMP occulter]} 
  \label{fig:11Apr_DVLW}
 \end{figure*}
To highlight these travelling features, Fig.~\ref{fig:11Apr_DVLW} displays the variation of Doppler velocity and line width along the arc as a function of time. In Fig.~\ref{subfig:11Apr_DVTD} we see ridges of features predominantly travelling upward along the loops; evidence of similar features may be seen in line width (Fig.~\ref{subfig:11Apr_LWTD}). We also remove a $9\,\rm{min}$ running average from each signal, in order to further enhance these ridges. Varying the period over which the running average is taken does not significantly affect our results. Using this technique, Fig.~\ref{subfig:11Apr_DVTDRAs} now illustrates Doppler-shifted ridges which maintain the same relative amplitude along the entire loop (of $1\mathord{-}2\,\rm{km~s}^{-1}$). Almost all the ridges are inclined to the right (indicating motion predominantly outward from the limb); many do not remain at the same inclination for all time, suggesting that the speeds of individual features may be changing as they propagate along the loop. The detrended line width image, Fig.~\ref{subfig:11Apr_LWTDRAs}, also contains some ridge-like features. The amplitude of these ridges are small ($400\mathord{-}500\,\rm{m\,s}^{-1}$) close to the base of the fan loop, but increase with height (up to a peak of ${\mathord{\sim}}1\,\rm{km~s}^{-1}$ near the top of the arc).

   \begin{figure*}[t]
    \centering\capstart
    \subfloat[Outward (prograde) filtered Doppler]{\label{subfig:11Apr_proDV}\resizebox{0.95\textwidth}{!}{\includegraphics{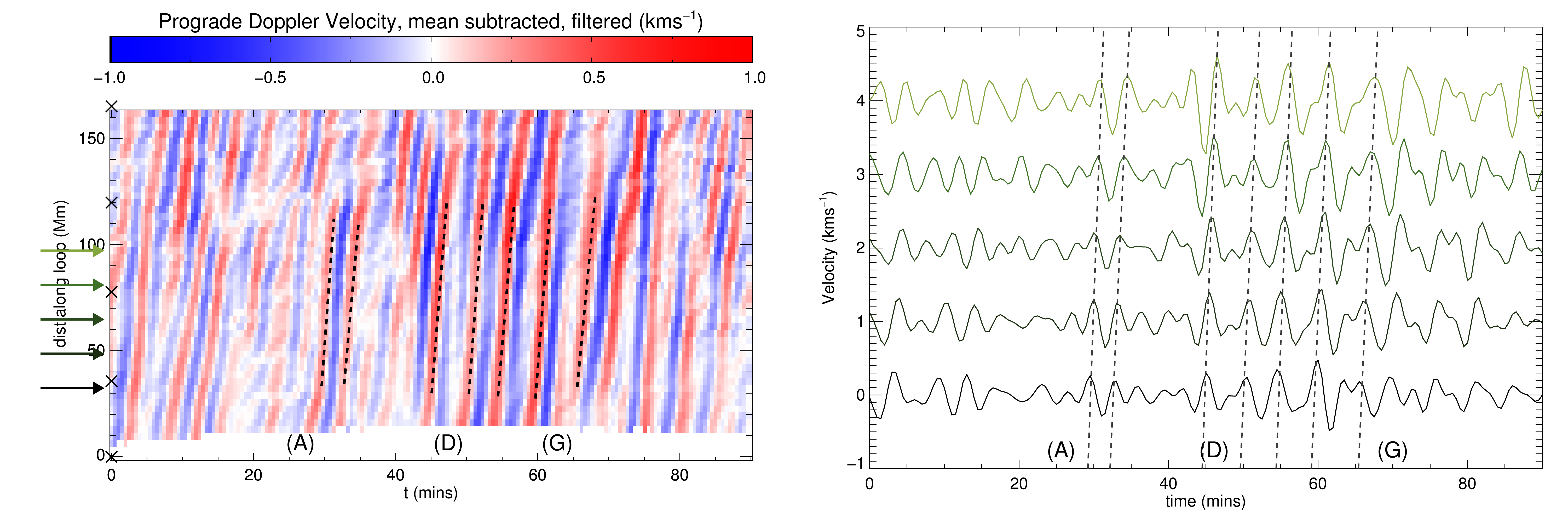}}}\\
    \subfloat[Inward (retrograde) filtered Doppler]{\label{subfig:11Apr_retDV}\resizebox{0.95\textwidth}{!}{\includegraphics{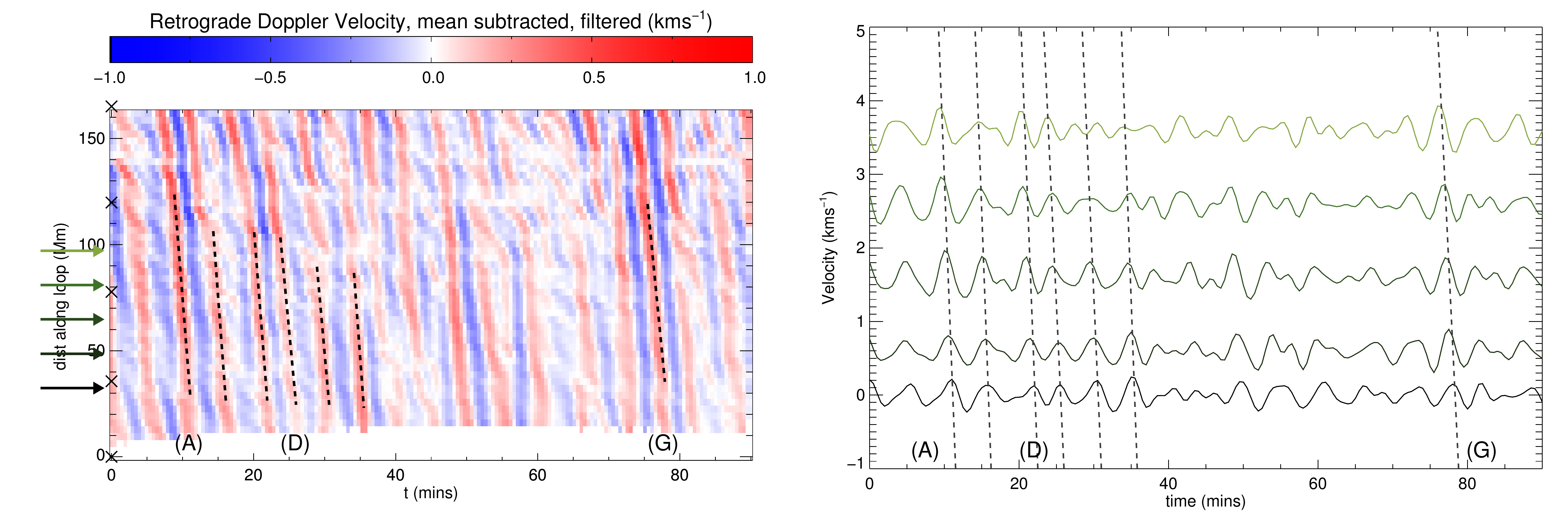}}}
    \caption{Reconstruction (using Fourier analysis) of maps and cross sections of Doppler motion along the loop, for periods of $3-8\,\rm{mins}$; \protect\subref{subfig:11Apr_proDV} shows prograde while \protect\subref{subfig:11Apr_retDV} shows retrograde features, whose speeds are estimated using the dashed black lines.}
    \label{fig:11Apr_proret}
   \end{figure*}

Following \citet{paper:TomczykMcIntosh2009}, we separate these features by their direction of propagation, focussing on waves in the $3\mathord{-}8\,\rm{min}$ range of periods ($5.5\mathord{-}2.1\,\rm{mHz}$). Figure~\ref{subfig:11Apr_proDV} illustrates prograde motion, i.e. Doppler-shifts which travel outward from the base of the visible loops. Ridges inclined in the opposite direction are shown in Fig.~\ref{subfig:11Apr_retDV}, representing Doppler motions propagating down towards the limb, which we label as retrograde motion. 

Splitting the diagrams in this way reduces the amplitudes of the ridges in Fig.~\ref{fig:11Apr_DVLW}. Ridges in the prograde image (Fig.~\ref{subfig:11Apr_proDV}) typically have amplitudes of $500\,\rm{m\,s}^{-1}$, with a maximum of around $700\,\rm{m\,s}^{-1}$. Retrograde amplitudes are slightly lower, typically around $300\,\rm{m\,s}^{-1}$, and peaking at approximately $500\,\rm{m\,s}^{-1}$.

Estimates of the propagation speeds of these waves are obtained from the gradients of the diagonal bands in Fig.~\ref{fig:11Apr_proret}. The average gradient of the prograde bands in Fig.~\ref{subfig:11Apr_proDV} is $692\,\rm{km~s}^{-1}$ (in a range from $591\mathord{-}765\,\rm{km~s}^{-1}$). The average gradient of retrograde bands in Fig.~\ref{subfig:11Apr_retDV} is $677\,\rm{km~s}^{-1}$ (from $572\mathord{-}762\,\rm{km~s}^{-1}$). These values are checked against features present in equally spaced cross sections of the space-time diagram. It should be noted that this method assumes that the speeds are constant. In order to verify these phase speed estimates, we also used a cross-correlation technique \citep[outlined in][]{paper:TomczykMcIntosh2009} to estimate a global average speed for each space-time diagram. This technique correlates slices through the diagram at different heights. The estimated lag time between each slice and the slice separation distance are combined to form an average phase speed. In the case of the prograde motions, a global average speed of $631\pm21\,\rm{km~s}^{-1}$ is recovered, while retrograde motions return a global average of $661\pm27\,\rm{km~s}^{-1}$ via the correlation method. A detailed discussion of techniques and errors for the recovery of speed information may be found in e.g. \cite{paper:Kiddieetal2012} or \cite{paper:YuanNakariakov2012}. The separation of each band was also used to estimate a period for these ridges. Prograde features in Fig.~\ref{subfig:11Apr_proDV} are approximately separated by $3-6\,\rm{mins}$, while retrograde features (Fig.~\ref{subfig:11Apr_retDV}) have separations of $4-6\,\rm{mins}$. An overview of speeds and periods is given in Table \ref{tab:vandp}.

  \begin{figure*}[t]
    \centering\capstart
    \subfloat[Prograde filtered line width]{\label{subfig:11Apr_proLW}\resizebox{0.95\textwidth}{!}{\includegraphics{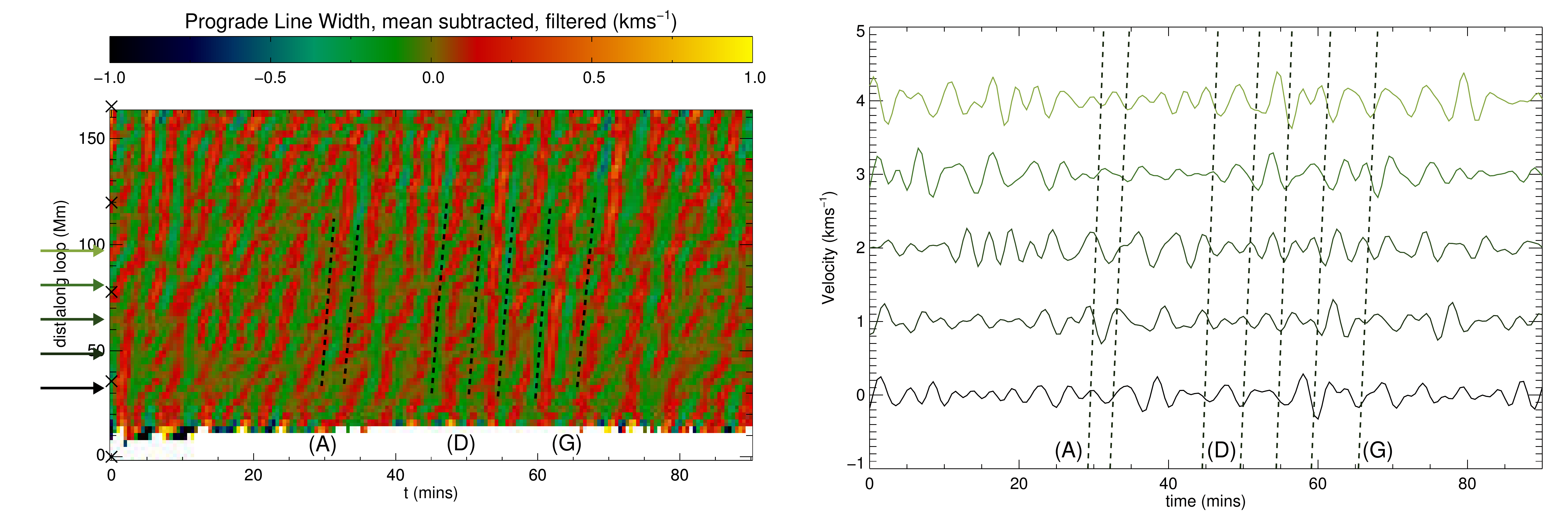}}}\\
    \subfloat[Retrograde filtered line width]{\label{subfig:11Apr_retLW}\resizebox{0.95\textwidth}{!}{\includegraphics{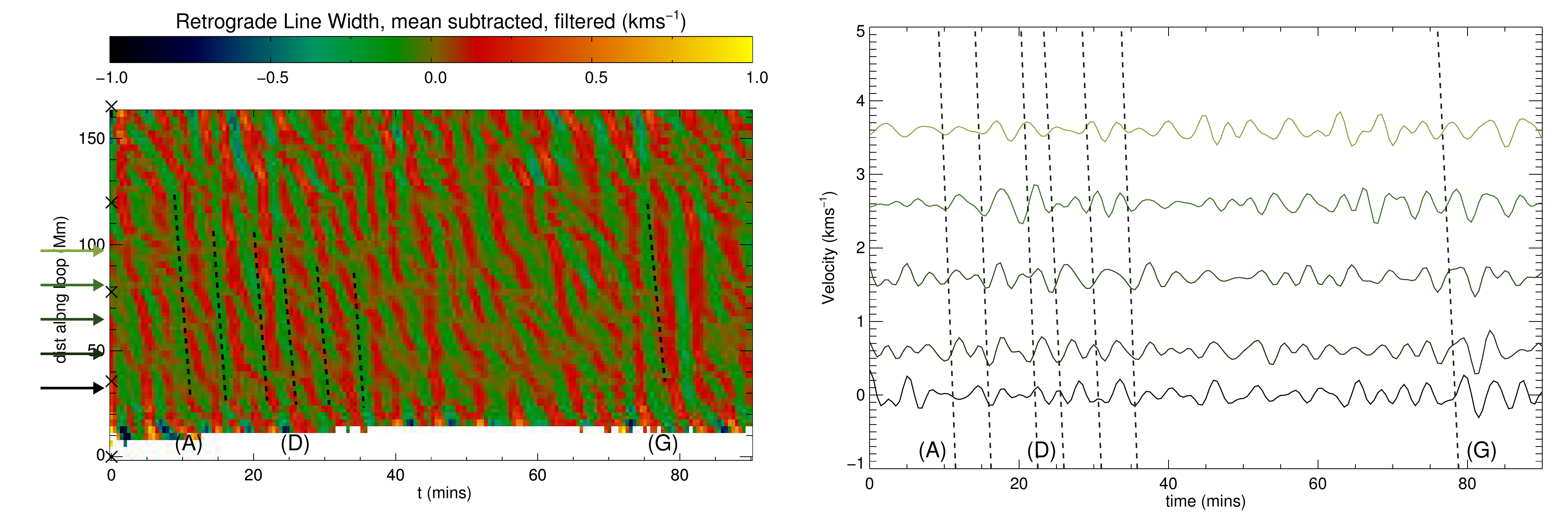}}}
    \caption{Reconstruction (using Fourier analysis) of maps and cross sections of line width variation along the loop, for periods of $3{\mathord{-}}8\,\rm{mins}$; \protect\subref{subfig:11Apr_proLW} shows prograde while \protect\subref{subfig:11Apr_retLW} shows retrograde features. The overlaid dashed lines are identical to those seen in Fig.~\ref{fig:11Apr_proret}.}
    \label{fig:11Apr_proretLW}
   \end{figure*}

Pro/retrograde filtering was also applied to the line width data seen in Fig.~\ref{subfig:11Apr_LWTD}; Fig.~\ref{subfig:11Apr_proLW} shows prograde filtered line width, while Fig.~\ref{subfig:11Apr_retLW} shows retrograde line width features, both filtered for $3{\mathord{-}}8\,\rm{min}$ periods. Typical prograde amplitudes are around $200\,\rm{m\,s}^{-1}$ with a maximum of $400\,\rm{m\,s}^{-1}$, while retrograde features are again slightly smaller, averaging $100\mathord{-}200\,\rm{m\,s}^{-1}$ up to a peak of $400\,\rm{m\,s}^{-1}$. The peak values are always observed close to the top of the track; almost all line width features appear to increase in amplitude with height. These variations appear to depend more strongly upon height above the limb than the Doppler variations seen in Fig.~\ref{fig:11Apr_proret}. A comparison of Fig.~\ref{fig:11Apr_proret} with Fig.~\ref{fig:11Apr_proretLW} shows similar ridge-like features in many of the same locations. The overplotted bands in Fig.~\ref{fig:11Apr_proretLW} are taken directly from the Doppler images. Many of these bands are aligned with features in the line width images.

\begin{table*}[t]
 	\caption{Phase speed and period estimates}
      \label{tab:vandp}
      \centering
        \begin{tabular}{lllrrrclrcl}\hline\hline
 Date&Instrument&Fig.&Oscillation&\multicolumn{3}{ c }{Band speed ($\rm{km~s}^{-1}$)}&Avg. speed     &\multicolumn{3}{ c }{Band separation (mins)}	\\
     &      &    &type       & max.     & min.        &avg. 		  	&(coh. method)  &max.    &min. 	&avg.   				\\
               \hline
 2012 Apr. 11&CoMP Doppler (prograde)&\ref{subfig:11Apr_proDV}&Transverse&$765$ &$591$ &$692$&$631\pm21$&$6$&$3$&$4.7$  \\ 
 2012 Apr. 11&CoMP Doppler (retrograde)&\ref{subfig:11Apr_retDV}&Transverse&$762$ &$572$ &$677$&$661\pm27$&$6$&$4$&$4.9$   \\
 2012 Apr. 11&AIA $193\angstrom$&\ref{fig:pcds11}&Longitudinal&$252$&$155$ &$190$&$-$&$11$&$8$&$9.5$   \\
 2012 Apr. 14&AIA $193\angstrom$&\ref{fig:pcds14}&Longitudinal&$193$&$105$&$139$&$-$&$10$&$6.5$&$8$\\
               \hline
           \end{tabular}
\tablefoot{Summary of speed and period information obtained from space-time analysis of fan loops anchored in the same active region, observed on 11 and 14 April 2012.} 
 \end{table*}

\section{11 April 2012: AIA}\label{sec:AIA11}

We now turn our attention to AIA observations of the same region on 11 April 2012. We use both parallel and perpendicular cuts along/across the arc in Fig.~\ref{fig:11Apr_refim} in order to compare wave-like behaviour with CoMP observations of the same structure (see discussion in Sec.~\ref{sec:meth}).

\subsection{Transverse motions}

\begin{figure*}[t]
   \centering\capstart
    \subfloat[Location A]{\label{subfig:11Apr_wobA}\resizebox{0.32\textwidth}{!}{\includegraphics{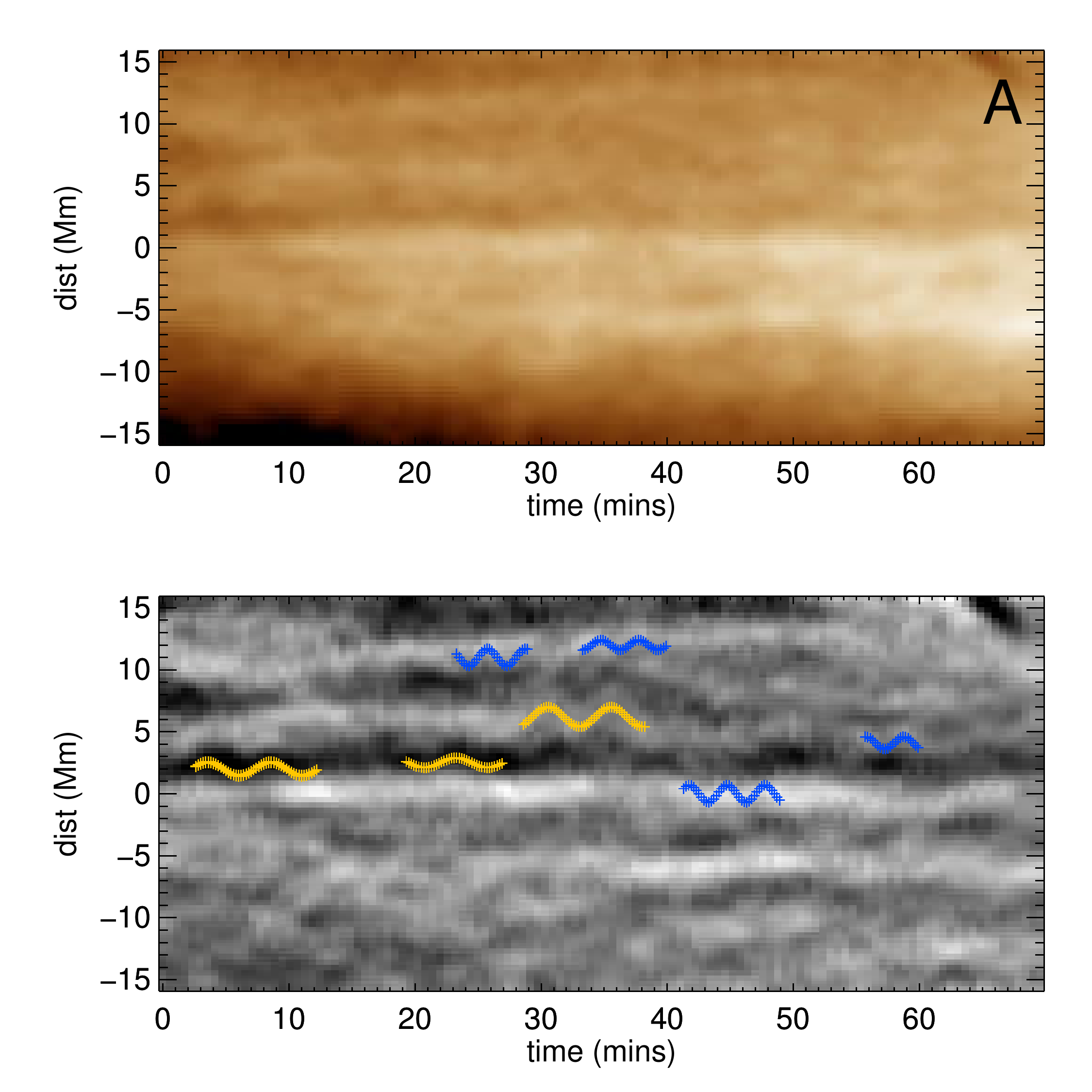}}}
    \subfloat[Location B]{\label{subfig:11Apr_wobB}\resizebox{0.32\textwidth}{!}{\includegraphics{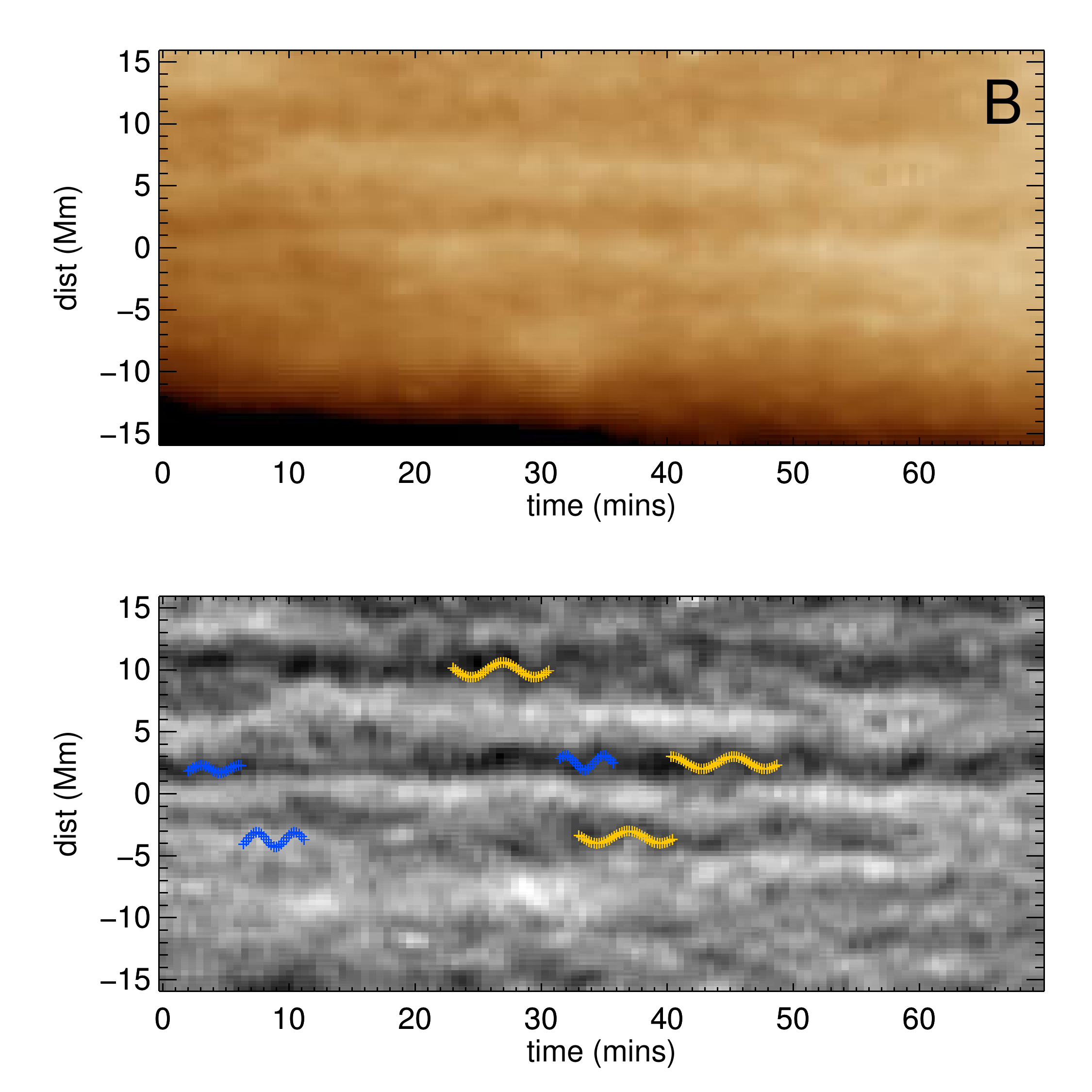}}} 
    \subfloat[Location C]{\label{subfig:11Apr_wobC}\resizebox{0.32\textwidth}{!}{\includegraphics{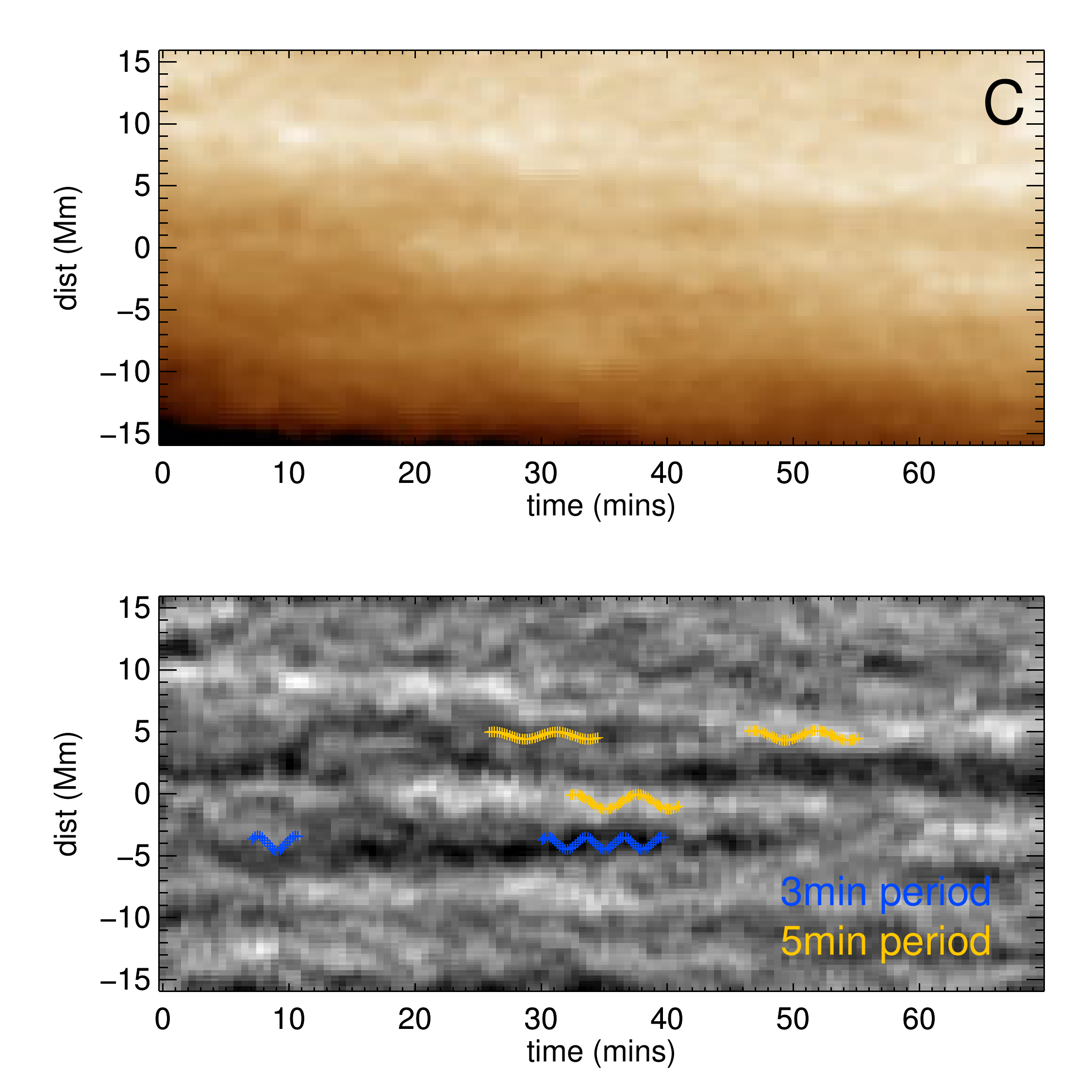}}} 
  \caption{Transverse motions, obtained using perpendicular cuts through the chosen track on 11 April 2012, for the positions seen in Fig.~\ref{subfig:limbAIAUSM}. Each frame of $193\angstrom$ data is unsharp masked, before both are sampled at the same cut location. The top row shows the variation of pure intensity, while the bottom row shows the corresponding unsharp mask image. Several features with periods close to $3$ and $5\,\rm{mins}$ have been highlighted in blue and yellow.} 
   \label{fig:11thwobbles}
  \end{figure*}

We use three perpendicular cuts (located as shown in Fig.~\ref{subfig:limbAIAUSM}) to investigate whether transverse motions observed along the line-of-sight in CoMP Doppler velocity are accompanied by loop-displacements in the plane-of-sky (see Fig.~\ref{fig:cartoon}). The space-time diagrams generated by these cuts are seen in Fig.~\ref{fig:11thwobbles}.

Due to the rapid fall-off in intensity with height, and the extreme line-of-sight superposition of features at the limb, identification of potential oscillatory features in images is not trivial; this is seen in the top row of Fig.~\ref{fig:11thwobbles}, which shows the variation of AIA $193\angstrom$ intensity at each cut in time. Applying an unsharp mask filter to each frame of AIA data, before repeating the sampling process at the same locations boosts the visibility of signatures of oscillatory motion; the results of this process are seen in the bottom row of Fig.~\ref{fig:11thwobbles}. In the unsharp mask space-time plots, several features demonstrate periods within the $100-500\,\rm{s}$ range as observed by \citet{paper:McIntoshetal2011}. To visualise this, we overlay specific features with an artificial sinusoidal signal, with a period of either $3$ or $5\,\rm{mins}$. This simple illustrative tool highlights the periodicities of the features in Fig.~\ref{fig:11thwobbles}. These short-period features rarely last more than $1-2$ wave periods, and typically displace loops by $<1\,\rm{Mm}$ (the overlaid sinusoidal trends have amplitudes from $0.3-0.8\,\rm{Mm}$). Wave amplitudes remain approximately constant over the lifetime of each oscillation, suggesting that little damping takes place.

\subsection{Longitudinal motions}

\begin{figure}[t]
    \centering\capstart
     \resizebox{0.48\textwidth}{!}{\includegraphics{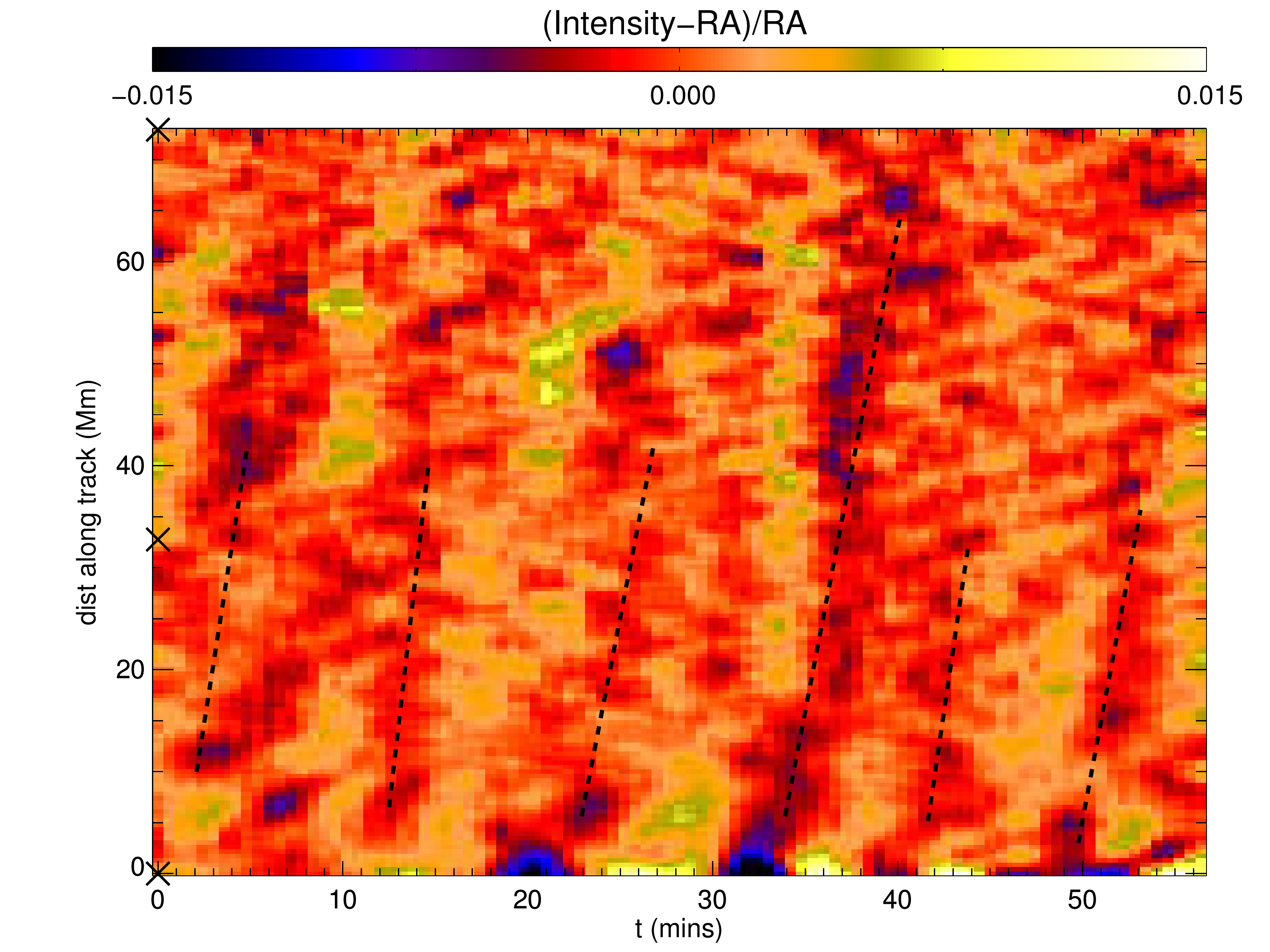}}
  \caption{Quasi-periodic longitudinal features along fan loops above the limb on 11 April 2012 in AIA $193\angstrom$. The dashed lines are used to estimate speeds and periods of the ridges; a $9\,\rm{min}$ running average has been used to establish amplitudes as a function of background intensity.} 
    \label{fig:pcds11}
    \end{figure}

An initial comparison of space-time images sampling intensity along the arc in AIA reveals several alternating dark and light ridges up to $40-50\,\rm{Mm}$ above the limb. In CoMP, this region is obscured by the occulter. Examples of these quasi-periodic longitudinal motions may be seen in Fig.~\ref{fig:pcds11} (where we estimate the intensity of these features as a fraction of the background by subtracting a $9\,\rm{min}$ running average).

Propagation speeds of $155-252\,\rm{km~s}^{-1}$ (with an average of $190\,\rm{km~s}^{-1}$) are recovered through a series of overplotted bands. We are unable to use the coherence based method to compute global average speeds due to low signal-to-noise. The separation of bands in Fig.~\ref{fig:pcds11} implies a much larger period than earlier transverse motions seen in CoMP Doppler velocity; features are separated from between $8-11\,\rm{mins}$, with an average of $9.5\,\rm{mins}$ (see Table \ref{tab:vandp}).

The amplitudes of these quasi-periodic intensity disturbances are extremely small, where the largest amplitudes are found close to the solar limb. We estimate amplitudes of between $1-1.5\%$ of the background intensity close the limb; above the limb the amplitude often falls below $0.5\%$. At these levels, distinguishing propagating disturbances from the background is extremely difficult.

\subsection{Coherence of longitudinal motions}

Many previous investigations which observe longitudinal quasi-periodic intensity disturbances use a summation across magnetic structures (as discussed in Sec.~\ref{sec:meth}). Figure~\ref{fig:pcds11} is created by summing over a distance of $1.76\,\rm{Mm}$ centred on the chosen arc (found by summing two pixels to the left and right, i.e. the five pixels at the centre of each perpendicular cut). Space-time images generated using just a single pixel track (using only the central pixel in each cut, for example) often contain a large amount of noise, which reduces the visibility of propagating features. Despite using high resolution images to assist in the track placement, a single pixel track is also unlikely to always map directly onto a single field-line/magnetic structure. Using a local summation not only boosts signal-to-noise, but also ensures that the magnetic structure of interest always contributes to the signal recovered along the summed track.

\begin{figure*}[t]
  \centering \capstart
    \resizebox{0.98\textwidth}{!}{\includegraphics{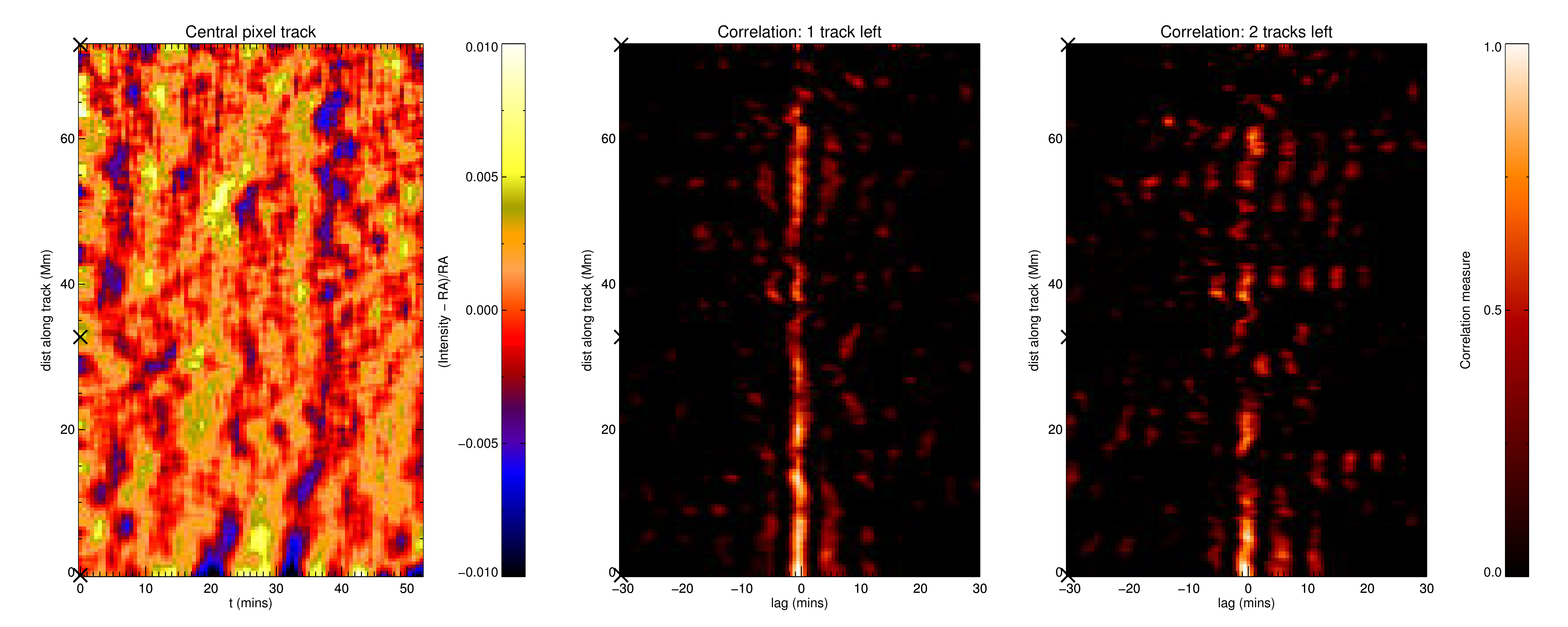}}
  \caption{Space-time plots, illustrating the correlation between individual neighbouring tracks along fan loops on 11 April 2012. The left hand image illustrates the space-time image obtained by the central pixel in each perpendicular cut, while the middle and right hand images imply the correlation of this with a similar image obtained along neighbouring tracks.}
  \label{fig:11thcoherence}
 \end{figure*}

{{We compare tracks formed by single pixels, at various positions in each perpendicular cut across the loop. This allows us to estimate how far we may move across a structure and still obtain related behaviour. The central track is used as a reference point.}} The space time image formed along this central track can be seen in the left hand panel of Fig.~\ref{fig:11thcoherence}. We then use a correlation routine to compare this with images obtained along neighbouring (single pixel) tracks which are progressively further away from the centre. In Fig.~\ref{fig:11thcoherence}, we obtain correlation plots from tracks one and two pixels to the left of centre, correlated against the central track. The output from the correlation routine is given as a function of lag along the horizontal axis; well correlated images would have strong peaks close to a lag of zero. Any peaks to the left/right of centre in the horizontal axis in correlation plots imply correlated behaviour between signals which have been phase-shifted.

The central panel of Fig.~\ref{fig:11thcoherence} shows a large amount of correlation between tracks from neighbouring pixels. The highest degree of correlation only exists close to the limb, and degrades with height; no correlated behaviour is observed above ${\mathord{\sim}}60\,\rm{Mm}$. Moving an additional pixel to the left of centre (right hand panel of Fig.~\ref{fig:11thcoherence}) reveals that only features very near the solar limb can be linked with any confidence. Few places above $10\,\rm{Mm}$ now achieve any correlation; those that do often also contain phase-shifted features showing similar coherence levels. In these cases the routine is no longer confident in relating a single feature in a given space-time image to a feature in the other space-time image, although a similar periodic signal still appears to be present. A track three pixels to the left of centre shows no coherent features. Similar results are recovered by progressively moving to the right of a central track; related behaviour is recovered on tracks up to a maximum of $2$ pixels to the left or right of centre. 

These findings allow us to estimate a coherence width for these observations. Two pixels to the left and right of centre are separated by ${\mathord{\sim}}1.8\,\rm{Mm}$. Assuming that the track at the centre contains the behaviour we wish to study, this range represents the maximum distance over which similar behaviour may be recovered, at low heights. Coherence is lost rapidly with increasing height \citep[c.f.][]{paper:Kingetal2003}.

\section{14 April 2012: AIA}\label{sec:AIA14}

\begin{figure}[t]
 \centering\capstart
 \resizebox{0.48\textwidth}{!}{\includegraphics{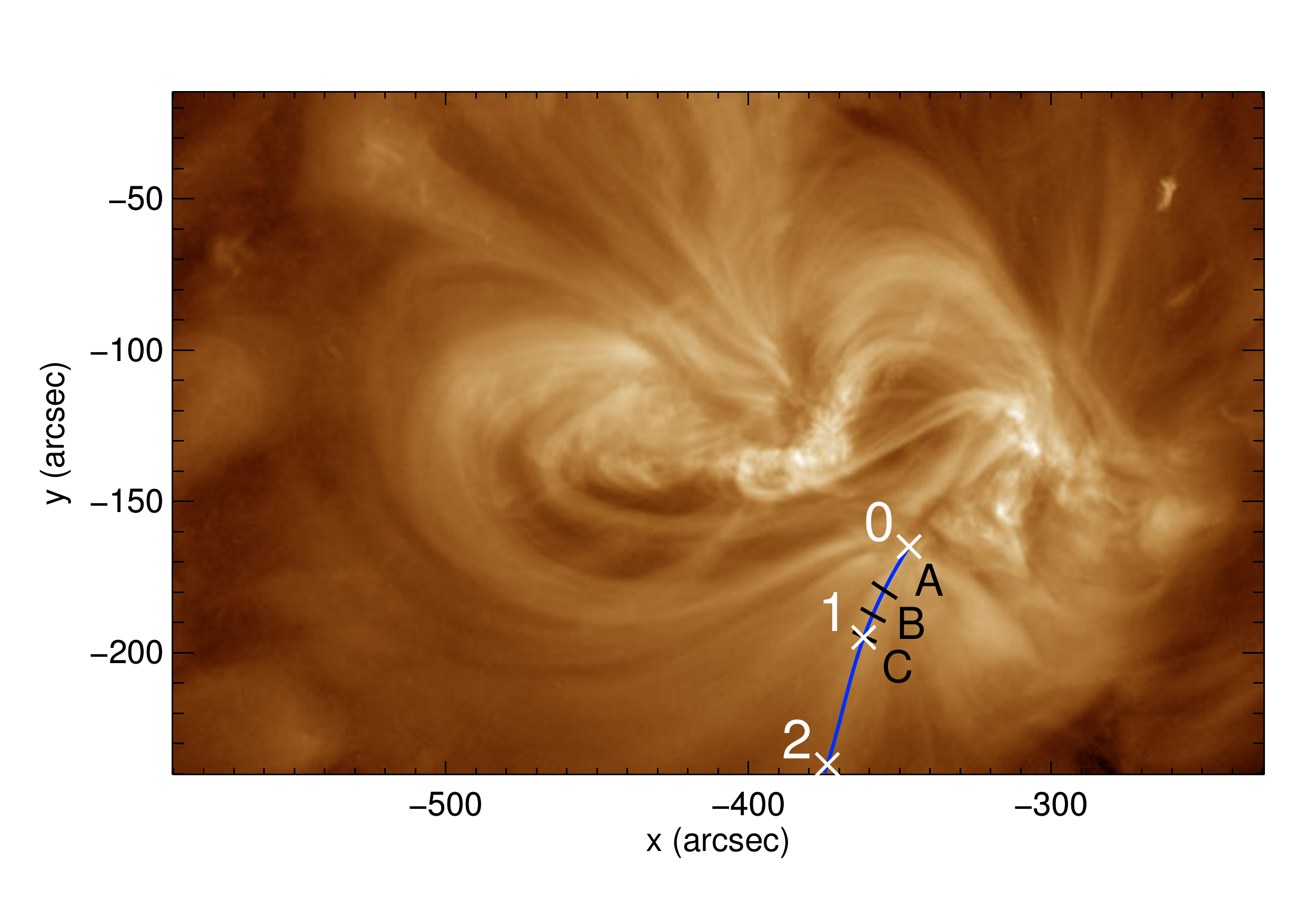}}
 \caption{Southern AR, tracked onto the disk, seen in AIA $193\angstrom$ intensity at 15:33UT on 14 April 2012, with fan loops anchored in the same region as those shown on the limb highlighted by the blue track. The location of several perpendicular cuts to this are also highlighted in black.}
 \label{fig:14Apr_refim}
\end{figure}

We return to the same active region at 15.33UT on 14 April 2012. Using images from STEREO-B and AIA, several fan-loops are visible, oriented in a similar direction and anchored in the same (approximate) location to those seen above the limb. We therefore chose to study one such fan-loop, and the specific track chosen is outlined in dark blue in Fig.~\ref{fig:14Apr_refim}.

\subsection{Transverse motions}

\begin{figure*}[t]
   \centering\capstart
    \subfloat[Location A]{\label{subfig:14Apr_wobA}\resizebox{0.32\textwidth}{!}{\includegraphics{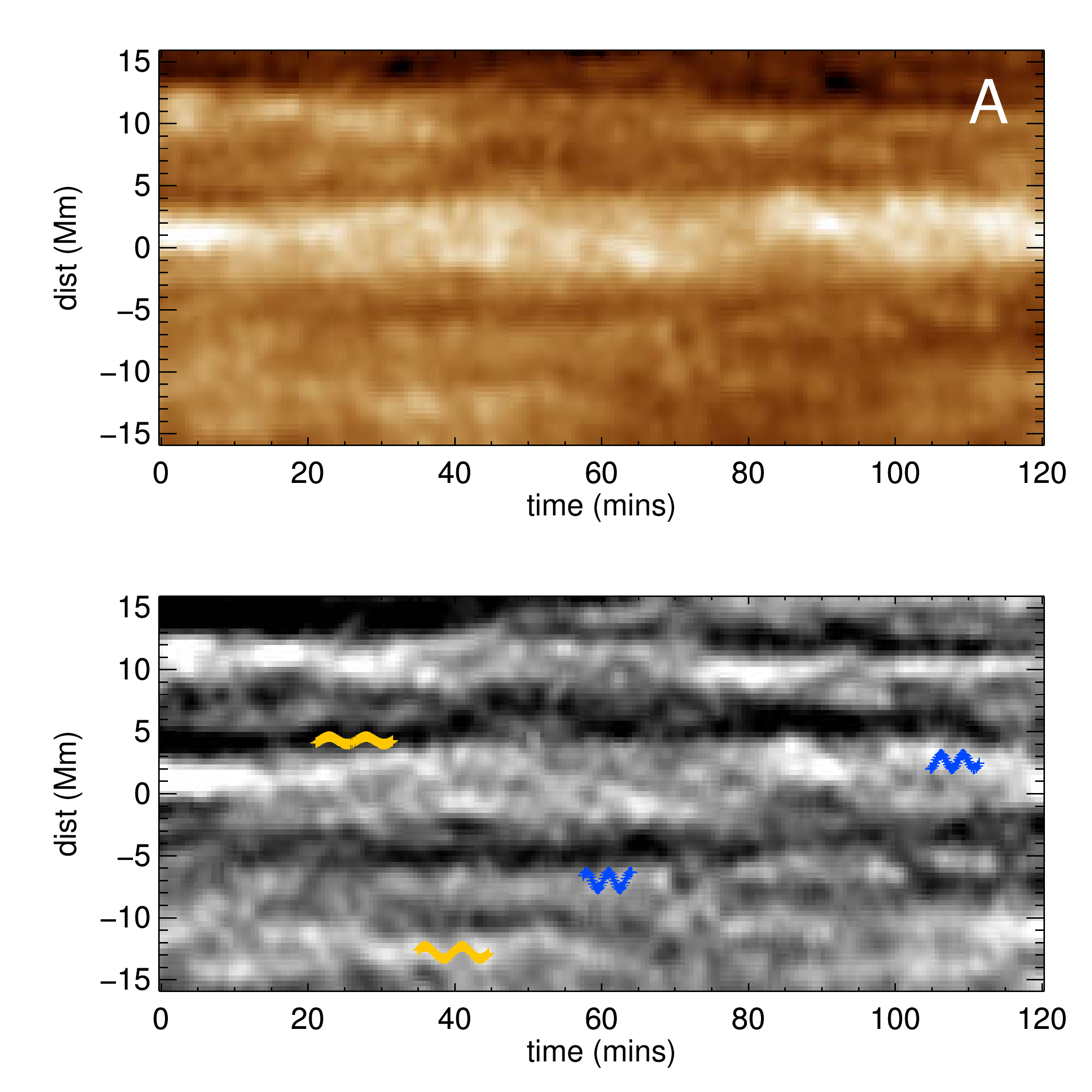}}}
    \subfloat[Location B]{\label{subfig:14Apr_wobB}\resizebox{0.32\textwidth}{!}{\includegraphics{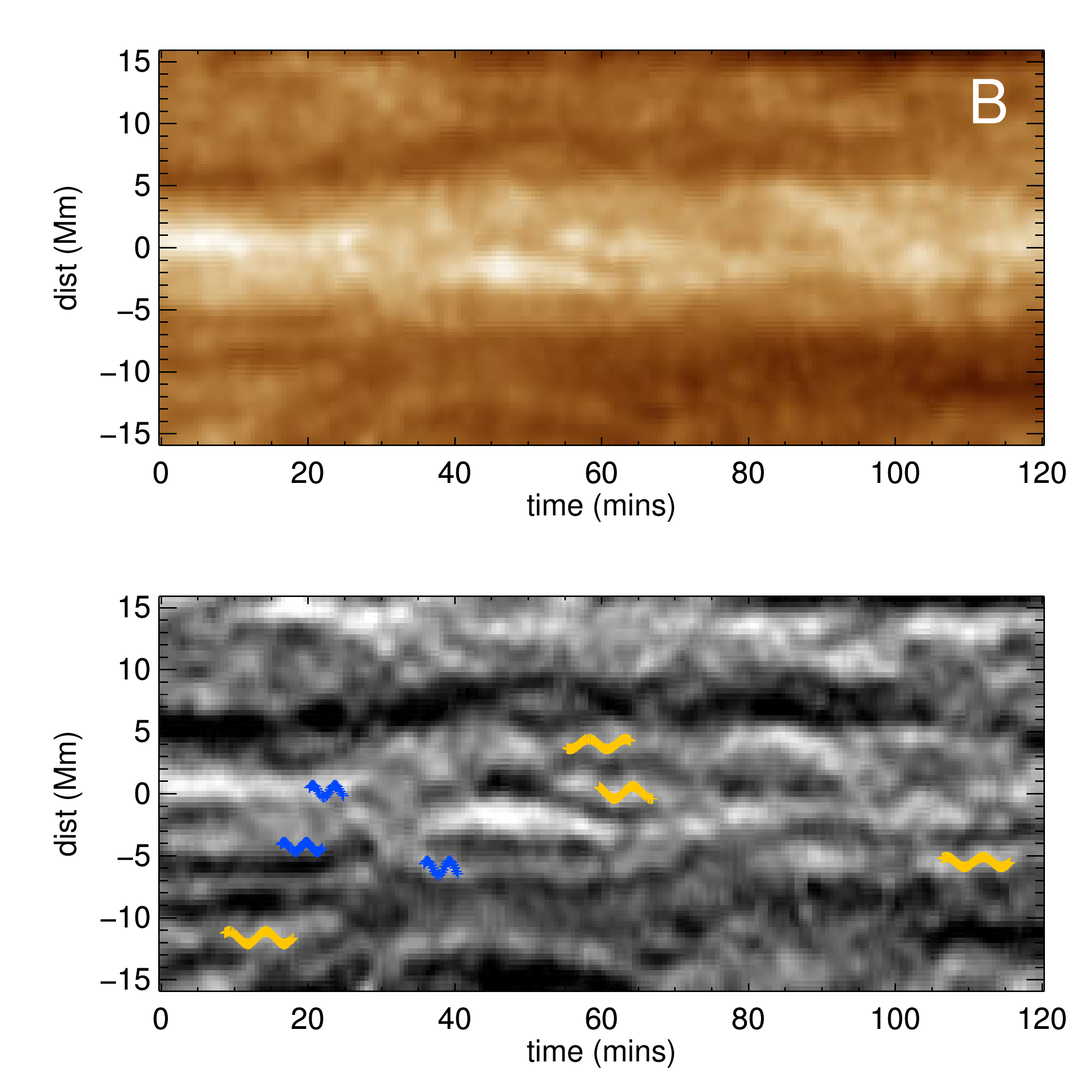}}} 
    \subfloat[Location C]{\label{subfig:14Apr_wobC}\resizebox{0.32\textwidth}{!}{\includegraphics{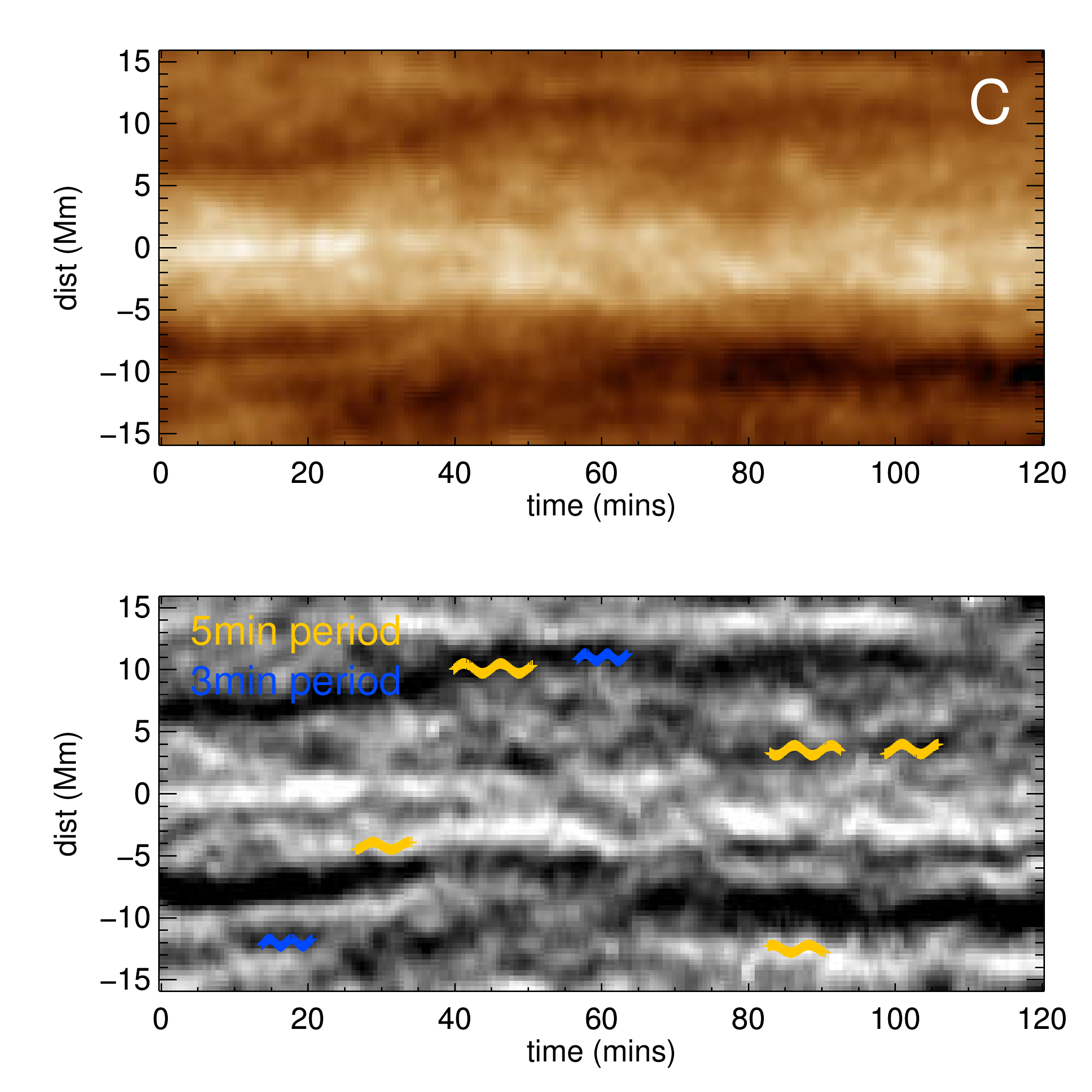}}} 
  \caption{Transverse motions, visible through perpendicular cuts across fan loop structures on 14 April 2012, at positions indicated in Fig.~\ref{fig:14Apr_refim}. {{Each $193\angstrom$ image is accompanied by an unsharp mask-sampled version;}} several features with periods close to $3$ and $5\,\rm{mins}$ have been highlighted in blue and yellow.} 
   \label{fig:14thwobbles}
  \end{figure*}

Without accompanying CoMP data (with the fan-loops fully on disk) we are only able to observe transverse loop displacements in the plane-of-sky using perpendicular cuts across loops in AIA. As before, three cuts are placed across our chosen loop, at the locations shown in Fig.~\ref{fig:14Apr_refim}. Data obtained from these perpendicular cuts are shown in Fig.~\ref{fig:14thwobbles}, where the presence of oscillatory features is again highlighted using artificial overlaid sinusoidal signals, with periods of both $3$ and $5\,\rm{minutes}$.

Images in the top panels of Fig.~\ref{fig:14thwobbles} show that the chosen loop is more readily identifiable in space-time diagrams, compared with those taken above the solar limb. The stronger contrast between the loop and its surroundings boosts the visibility of the many oscillatory features in the image. Unsharp-masked images (bottom panels of Fig.~\ref{fig:14thwobbles}) further enhance these oscillatory features. From this, several sub-structures within the main fan-loop structure can also be seen to oscillate. From the overlaid signals, it is clear that many of these disturbances have periods well within the $150-500\,\rm{s}$ range. As before, these features typically exist for $\lesssim2$ full wave-periods, with displacements ranging from $0.2-0.8\,\rm{Mm}$, and which remain relatively constant over the lifetime of the oscillation.

\subsection{Longitudinal motions}
\begin{figure}[t]
    \centering\capstart
     \resizebox{0.49\textwidth}{!}{\includegraphics{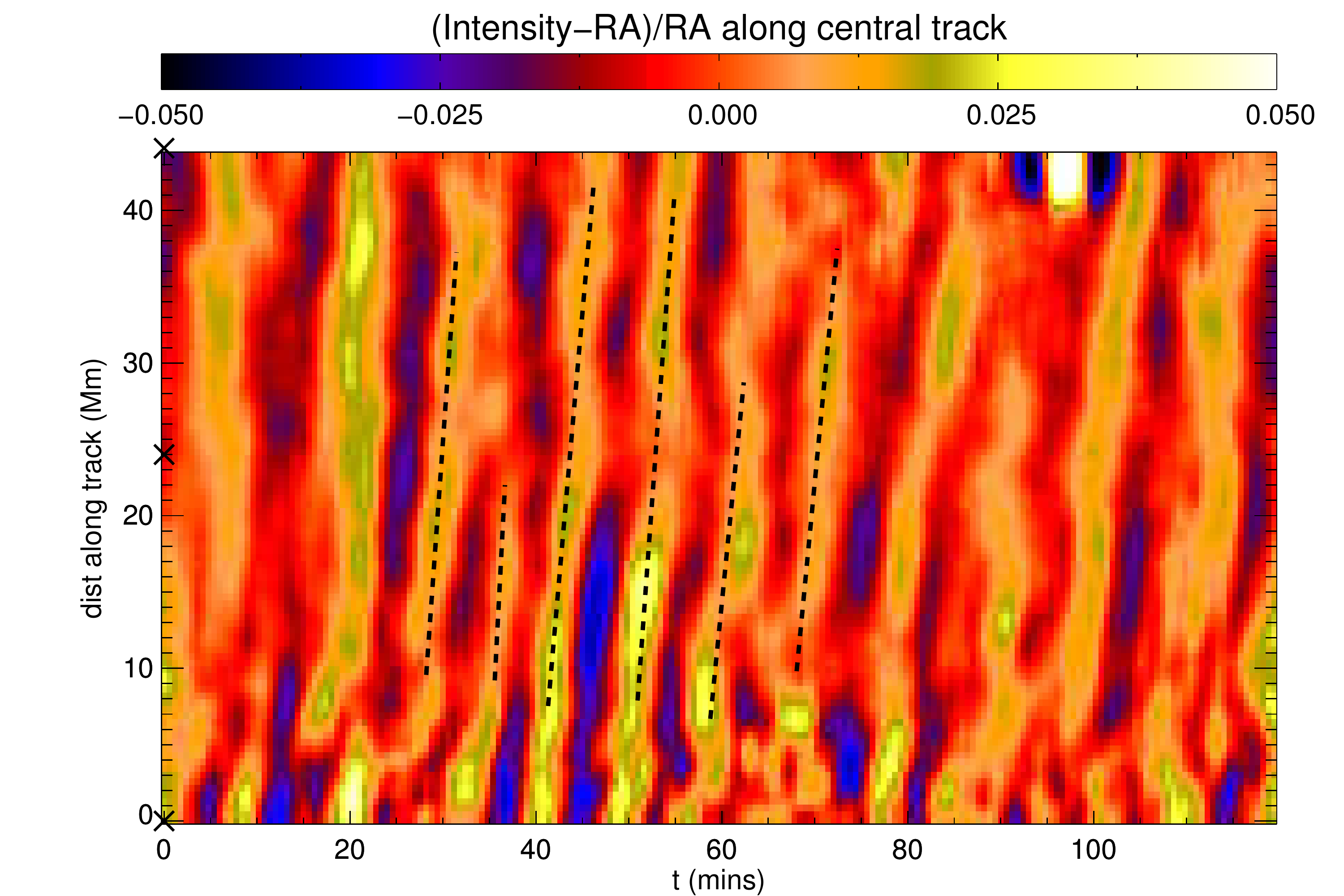}}
  \caption{Quasi-periodic features seen in space-time plots along on-disk fan loops on 14 April 2012 in $193\angstrom$. The gradients and separations of the overplotted dashed lines are used to estimate speeds and periods.} 
    \label{fig:pcds14}
   \end{figure}

Quasi-periodic propagating longitudinal intensity signatures are much clearer on 14 April 2012. Due to the comparatively high signal-to-noise, and the strong convergence of the loops upon the same footpoint, we will analyse intensity signals recorded along a single pixel track along the arc. In doing so, we hope to avoid summation across additional footpoints of loops which would no longer contribute to the signal once the loops begin to diverge apart.
Well defined ridges are clearly seen in the $193\angstrom$ space-time image of intensity along the arc (Fig.~\ref{fig:pcds14}). Speed estimates of these features (using the gradient of overplotted bands) range from $105-193\,\rm{km~s}^{-1}$, with peak-to-peak separations of $6-10\,\rm{mins}$. 
The amplitudes of features seen in Fig.~\ref{fig:pcds14} are much larger than those above the limb; though the amplitudes do not remain constant, they often range from $2-4\%$ of background. We also note the presence of higher frequency intensity perturbations close to the base of the tracks. Near to the loop footpoints, several features appear to only be separated by $\sim5\,\rm{mins}$. At the heights where we have begun to identify specific features, the dominant periods now appear to be closer to $7-10\,\rm{mins}$. 

Repeating the coherence analysis performed in Sec.~\ref{sec:AIA11} suggests that neighbouring parallel tracks close to the footpoints are likely to contain uncorrelated behaviour. Summation over parallel tracks should only be used in cases where loops do not diverge strongly.

\section{Discussion}\label{sec:disc}

Combining CoMP and SDO/AIA observations, we have found evidence of both longitudinal and transverse waves throughout the region of interest; off-limb observations by CoMP illustrate many displacements along the line-of-sight in Doppler velocity which travel along magnetic structures in the corona and are unaccompanied by significant intensity variations along the structure. These displacements are visible above the background Doppler velocity, which falls from approximately $0$ to $-5\,\rm{km\,s}^{-1}$ along the studied track (shown in Fig.~\ref{subfig:11Apr_DVTD}). Similar to the results of \citet{paper:Tomczyketal2007} and \citet{paper:TomczykMcIntosh2009}, these waves travel at speeds well above the local sound speed ($600-750\,\rm{km\,s}^{-1}$) with periods in the $3-6\,\rm{min}$ range,  along a single fan-loop. The amplitudes of these Doppler features seems somewhat larger than previous investigations. Whereas \citet[][]{paper:TomczykMcIntosh2009} found amplitudes of the order of $300\,\rm{m\,s}^{-1}$, we find values which vary from $300-700\,\rm{m\,s}^{-1}$. 

Following \citet{paper:TomczykMcIntosh2009}, we estimate the energy flux as
\[
 F_{\rm{trans}}=\rho\langle v^2\rangle v_p,
\]
where $\rho$ is the density, $v$ is the velocity amplitude and $V_p$ the phase speed. Using the same mass densities as \citet{paper:TomczykMcIntosh2009}, $2\times10^{-16}-2\times10^{-15}\rm{g\,cm^{-3}}$, together with our upper bounds of speed ($v_p=750\,\rm{km\,s}^{-1}$) and amplitude ($v=700\,\rm{m\,s}^{-1}$), we obtain a maximum energy flux carried by these waves of $75\mathord{-}750\,\rm{erg\,cm^{-2}s^{-1}}$. Despite this being a substantial increase on previous estimates, the energy flux falls far short of the $3\times10^{5}\,\rm{erg\,cm^{-2}s^{-1}}$ required to balance quiet solar coronal radiative losses \citep{paper:WithbroeNoyes1977}.

Our estimate of the energy flux is affected by several factors. As discussed in  \citet{paper:TomczykMcIntosh2009}, the above formula implies an interpretation of the observed transverse waves as (bulk) Alfv\'enic waves. An interpretation in terms of fast magneto acoustic kink waves \citep[e.g.][]{paper:VanDoorsselaereetal2008} or surface Alfv\'en waves \citep{paper:TGVinprep2013} could reduce the energy estimate by as much as an order of magnitude. On the other hand, the low spatial and temporal resolution of CoMP together with the large (off-limb) superposition along the line-of-sight will lead to a substantial underestimate of the energy content of the {\it resolved} Doppler oscillations. Both numerical simulations \citep{paper:DeMoortelPascoe2012} and a combination of observational analysis and Monte-Carlo simulations \citep{paper:McIntoshDePontieu2012} have demonstrated that this superposition of unresolved loop threads could lead to an underestimate of the true energy flux by as much as one to two orders of magnitude.

The CoMP Doppler-shifts have amplitudes which remain relatively constant over the length of the track. For isothermal loops at $T=1.6\,\rm{MK}$, the pressure scale height is approximately $80\,\rm{Mm}$ \citep[][]{book:Priest1982}; this distance is equivalent to half the length of the chosen track, suggesting that gravitational stratification will be important for waves propagating along this structure. \citet{paper:Soleretal2011} demonstrate the effect of longitudinal stratification on the propagation of resonantly damped MA kink waves using an analytical model. They find that both the wavelength and amplitude of the studied waves are affected by density variations along the loop. In particular, they note that resonant absorption and longitudinal stratification have opposite effects on the kink wave amplitude; while transverse density stratification typically leads to a reduction in wave amplitude due to mode-conversion to {\A} waves, longitudinal density variations can lead to wave amplitudes which increase with height, even while undergoing damping through mode coupling. Our observations may therefore contain evidence of two competing effects; in combination, longitudinal and transverse stratification can yield wave amplitudes which remain constant as a function of height. Similar studies have also been carried out for phase-mixed {\A} waves \citep{paper:DeMoorteletal1999} and for propagating slow waves \citep{paper:DeMoortelHood2004}.

Variations in CoMP Doppler velocity are also accompanied by variations in line width. Overall, the background emission line width increases from $40$ to $43\,\rm{km\,s}^{-1}$ over the length of the track (see Fig.~\ref{subfig:11Apr_LWTD}). Small fluctuations of this value also appear to grow with height, from typically $100-200\,\rm{m\,s}^{-1}$ near the base of the loop, to $400\,\rm{m\,s}^{-1}$ near the top of our track. The maximum amplitude of these variations approaches the size of features seen in CoMP Doppler and the propagation speeds (measured from the gradients in the time-distance diagrams) are of the same order as those found for the Doppler velocity perturbations.

We also observe both an overall broadening of emission lines with height, together with small periodic perturbations, whose amplitudes also increase with height. Several types of unresolved motions could be contributing to the observed non-thermal line broadening. Mode coupling combined with phase mixing will lead to an increase (with height) of unresolved, azimuthal ($m=1$) Alfv\'en waves. Torsional ($m=0$) Alfv\'en waves as reported by \citet{paper:DePontieuetal2012} will also lead to line broadening. The slight increase of the line broadening with height can again be explained in terms of the competing effects of wave damping and gravitational stratification (which would cause the amplitudes to increase with height). Estimates of wave speeds, periods and amplitudes would require higher resolution observations, both in space and time, such as those expected from, for example, IRIS.

Complementing the transverse loop displacements visible along the line of sight (seen in CoMP Doppler velocity), AIA results also show signatures of transverse motions in the plane of sky on the same fan-loop system. These signatures have periods in the $3-5\,\rm{min}$ range, and are visibly similar to features observed by \citet{paper:McIntoshetal2011}. Our results show displacements of the supporting loop structure of no more than $1\,\rm{Mm}$, and hence are unseen by CoMP intensity (whose pixel-width is ${\mathord{\sim}}3.24\,\rm{Mm}$), strengthening the conclusion of \citet{paper:McIntoshDePontieu2012} that the discrepancy between CoMP and AIA observations of propagating transverse waves can be explained by the lower spatial (and temporal) resolution of CoMP. {{Our results further add to the recent observational findings that oscillatory motions are ubiquitous throughout the solar atmosphere (e.g. as in \citealp{paper:Mortonetal2012}, \citealp{paper:DePontieuetal2012} for the chromosphere and \citealp{paper:McIntoshetal2011} in the corona).}}

Due to the fact that these are likely to be signatures of propagating waves, it should be possible to estimate the speed at which these waves travel, by identifying the time-lag between oscillatory features found in successive perpendicular cuts. There are several difficulties with this technique. Identifying the same oscillation using successive cuts is extremely difficult on the disk, and impossible above the limb, due to noise, LOS superposition and interference from neighbouring loop structures. We are also hampered by our temporal resolution; a series of propagating transverse displacements travelling at, say, $500\,\rm{km\,s}^{-1}$ would be seen in neighbouring perpendicular cuts (separated by $0.44\,\rm{Mm}$) only $0.9\,\rm{s}$ apart. Given our cadence of $36\,\rm{s}$, only cuts separated by $18\,\rm{Mm}$ would produce signals lagging by a single pixel in space-time plots. In order to yield a significant lag between signals (from which we might infer a speed), significant distances (${\mathord{\sim}}{50\,\rm{Mm}}$) are required. At present we are only able to positively identify the same oscillatory feature on two or three neighbouring perpendicular cuts (at best). Imaging data with a much higher cadence are required to recover a speed from signals which display such short lag times (${\mathord{\sim}}1{\,\rm{s}}$) between successive perpendicular cuts.
 
Our observations of longitudinal quasi-periodic disturbances qualitatively agree with previous observations of similar disturbances. Both the speeds ($100-250\,\rm{km\,s}^{-1}$) and periods ($6-11\,\rm{mins}$) are within the ranges summarised in, for example, \citet{review:DeMoortel2009} and \citet{paper:Banerjeeetal2011} for observations of propagating quasi-periodic longitudinal intensity features. The recovered amplitudes are relatively close to the background intensity level, particularly in observations above the limb. The signal-to-noise above the limb controls the width over which we observe similar behaviour along parallel structures. On disk, the strong divergence of the fan-loops means that the structures are no longer parallel. Typically, for strongly diverging fields, a common tool is to sum over a series of arcs, which all start at the same place; this compensates for the divergence of the field. Examples of this can be seen in, for example, \citet{review:DeMoortel2009}. 

We also note that the frequency of these disturbances appears to increase close to the loop footpoint. Similar period reductions are visible in previous AIA observations \citep[see e.g. Fig.~3 of][]{paper:McIntoshetal2012}. This reduction in the periods could either be apparent, for example due to the higher superposition near the footpoints caused by the diverging geometry or could be the consequence of a damping mechanism with an inherent period-dependency. For example, damping due to thermal conduction as suggested by \citet{paper:DeMoortelHood2003} will lead to longer damping lengths for perturbations with longer periods \citep[see also][]{paper:KrishnaPrasadetal2012a}. {{This would also explain the discrepancy between the larger periods recovered above the limb and the somewhat shorter periods found in the on-disk observations. The loop-parts observed above the limb are much further from their footpoints than their on-disk counterparts; damping of the shorter period features due to thermal conduction would mean only long period features are likely to be recovered at these heights. All the observed periods remain within the envelope of p-mode oscillations \citep[see e.g. Fig. 2 of][]{paper:TomczykMcIntosh2009}.

Following \citet{paper:McEwanDeMoortel2006}, we can estimate the energy flux of these longitudinal motions as
\[
 F_{\rm{long}}=\rho\left[\frac{\left(\delta v\right)^2}{2}\right] v_s,
\]
where $\delta v$ is the wave velocity amplitude, and $v_s$ is a measure of the local sound speed ($v_s\approx c_s=192\,\rm{km\,s}^{-1}$). Retaining the mass densities used to calculate the energy flux of transverse waves ($F_{\rm{trans}}$) yields estimates of the longitudinal wave energy flux, which range from $43\mathord{-}430\,\rm{erg\,cm^{-2}s^{-1}}$ above the solar limb (where intensity variations peak at $\sim1\%$) to $690\mathord{-}6900\,\rm{erg\,cm^{-2}s^{-1}}$ for motions observed on disk (peak intensity amplitude $\sim4\%$). Thus longitudinal motions potentially carry as much, if not more energy than their transverse counterparts, but still remain well below the amount of flux required to balance even Quiet Sun (coronal) radiative losses.}}

\section{Conclusions}\label{sec:conc}

Combining CoMP and AIA, our results clearly show that different wave motions are visible along the same structures: both (fast) transverse and (slower) longitudinal perturbations appear to be supported by the same structures.  Furthermore, combining line-of-sight Doppler velocity observations from CoMP and plane-of-sky intensity observations from SDO/AIA allows us to see two different components of the propagating transverse motions. Such a combination allows for the study of wave properties which cannot be studied using only a single (current) instrument; indeed, even observing these motions above the limb using imaging instruments is difficult. CoMP Doppler velocity oscillations travel along our fan loop at $600\mathord{-}750\,\rm{km~s}^{-1}$, well above the local sound speed. Sub-$\rm{Mm}$ displacements are difficult to observe with CoMP, but are resolved by AIA. In tandem with LOS Doppler motions, we also see evidence of periodic line width features in CoMP, whose amplitudes grow with height. Both findings support theoretical and numerical models which suggest that these waves exist as a coupled kink/{\A} mode. 

Much closer to the loop footpoints, we also see evidence of longitudinal propagating intensity features along the same structure. The properties of these motions (speeds of $100-200\,\rm{km\,s}^{-1}$, periods of $5-11\,\rm{mins}$, small amplitudes $<5\%$ of background) agree well with previous studies of similar features. The range over which these longitudinal motions are reported also contains examples of transverse oscillatory features. How these different types of waves exist on the same structure at the same time and how they interact are important questions, both in a seismological context and in order to establish the role each plays in the deposition of energy and mass to the corona. In short, interpretation of these waves in terms of one ``mode'' along the entirety of an isolated coronal structure is difficult, to say the least. The next generation of instruments (e.g.~IRIS or COSMO/ChroMag) will be required to spectrally sample a large enough field of view, in a very short time, to accurately diagnose simultaneous (combined) longitudinal and transverse wave properties.

\begin{acknowledgements}
The authors would like to thank S. Tomczyk for helpful discussions regarding the interpretation of CoMP data. JT would like to thank D. Pascoe, for assistance with Fig.~\ref{fig:cartoon}, and G. Kiddie, for useful discussions. IDM acknowledges support of a Royal Society University Research Fellowship. The research leading to these results has also received funding from the European Commission Seventh Framework Programme (FP7/2007-2013) under the grant agreements SOLSPANET (project No. 269299, \url{www.solspanet.eu/solspanet}).
\end{acknowledgements}

\bibliographystyle{aa}        
\bibliography{21782}          
\end{document}